\documentclass[notitlepage, amsmath, nofootinbib, aps, reprint, superscriptaddress,floatfix]{revtex4-1}
\usepackage{float}

\usepackage{url}
\usepackage{color}
\usepackage{rotating}
\usepackage{amsmath}
\usepackage{tabu}
\usepackage{hyperref}
\usepackage{graphicx}
\usepackage{array,multirow}
\usepackage[export]{adjustbox}
\graphicspath{ {images/} }
\usepackage{tcolorbox}
\usepackage{diagbox}
\usepackage{colortbl}
\definecolor{redD}{rgb}{1,0.8,0.8}

\newcommand{\lcdm}{$\Lambda{\rm CDM}$}

\hypersetup{
    colorlinks=true,
    linkcolor=blue,
}

\begin{document}
\title{Misinterpreting Modified Gravity as Dark Energy: a Quantitative Study}
\date{\today}

\author{Yuewei Wen}
\email{ywwen@umich.edu } 
\affiliation{Department of Physics, University of Michigan, 450 Church St, Ann Arbor, MI 48109-1040}

\author{Eva Nesbit}
\email{ehnesbit@syr.edu} 
\affiliation{Department of Physics, Kenyon College, Gambier, Ohio 43022}
\affiliation{Department of Physics, Syracuse University, Syracuse, NY 13244}

\author{Dragan Huterer}
\email{huterer@umich.edu } 
\affiliation{Department of Physics, University of Michigan, 450 Church St, Ann Arbor, MI 48109-1040}
\affiliation{Leinweber Center for Theoretical Physics, University of Michigan, 450 Church St, Ann Arbor, MI 48109-1040}
\affiliation{Max-Planck-Institut f\"ur Astrophysik, Karl-Schwarzschild-Str.\ 1, 85748 Garching, Germany}

\author{Scott Watson}
\email{gswatson@syr.edu} 
\affiliation{Department of Physics, Syracuse University, Syracuse, NY 13244}

\begin{abstract}
Standard cosmological data analyses typically constrain simple phenomenological dark-energy parameters, for example the present-day value of the equation of state parameter, $w_0$, and its variation with scale factor, $w_a$. However, results from such an analysis cannot easily indicate the presence of modified gravity. Even if general relativity does not hold, experimental data could still be fit sufficiently well by a phenomenological $w_0w_a$CDM, unmodified-gravity model. Hence, it would be very useful to know if there are generic signatures of modified gravity in standard analyses. Here we present, for the first time to our knowledge, a quantitative mapping showing how modified gravity models look when (mis)interpreted within the standard unmodified-gravity analysis. Scanning through a broad space of modified-gravity (Horndeski) models, and assuming a near-future survey consisting of CMB, BAO, and SNIa observations, we report values of the best-fit set of cosmological parameters including $(w_0, w_a)$ that would be inferred if modified gravity were at work. We find that modified gravity models that can masquerade as standard gravity lead to very specific biases in standard-parameter spaces. We also comment on implications for measurements of the amplitude of mass fluctuations described by the parameter $S_8$.
\end{abstract}

\maketitle

\section{Introduction} \label{sec:introduction}

Overwhelming observational evidence for the current acceleration of the universe presents one of the most outstanding theoretical challenges in all of cosmology and physics \cite{Frieman:2008sn,Huterer:2017buf}. The physical mechanism for the apparent acceleration remains fundamentally mysterious. It could be given by the presence of the cosmological-constant term in Einstein's equations, but the tiny size of the constant presents an apparently insurmountable challenge \cite{Weinberg:1988cp,Carroll:2000fy}. A number of dark energy models beyond the cosmological constant have been proposed as well \cite{Copeland2006}. Similarly, the accelerated expansion could be that gravity is modified on large scales \cite{Clifton2012, Joyce2015, Silvestri2009}, but thus far there is no direct evidence for such a modification. 

The difficulty with studying modified-gravity models with data is that the space of possibilities is enormous. There are many completely distinct classes of models to modify gravity and, in each, a large number of possible parameterizations. Constraining any \textit{one} of those modified-gravity model parameterizations with large-scale structure also presents a challenge, for the following reasons: i) modified-gravity-model predictions for nonlinear clustering are, with a few exceptions, not available at all; and ii)  the linear-theory predictions generally need to be validated by (modified-gravity) N-body simulations, as e.g.\ galaxy bias in these models may differ from that in standard gravity (for example \cite{Arnold2019, Mitchell:2021uzh}). Tests of modified gravity with the cosmic microwave background (CMB) are a little easier as one only needs linear-theory predictions and there is no galaxy bias, but the large scale of possible modified-gravity theories still presents a major obstacle.

As a consequence of these challenges, the majority of confrontations of theory with data has not encompassed models of modified gravity. Instead, most analyses consider simple phenomenological descriptions of the dark-energy sector, such as the model with a cosmological constant (\lcdm), and that with constant dark-energy equation of state parameter $w$ (wCDM) \cite{Turner:1997npq}. Also popular is the time-varying parameterization of the dark-energy equation of state \cite{Linder2003} that allows for the dynamics, $w(a) = w_0 + w_a(1-a)$, where $a$ is the scale factor and $w_0$ and $w_a$ parameters to be constrained by the data. Modified gravity has typically been constrained only for very specific models (e.g.\ $\Sigma, \mu$ parameterizations of the gravitational potentials, \cite{Zhang:2007nk,Daniel:2012kn,Pogosian:2016pwr}). There have been attempts to constrain individual modified-gravity models \cite{Zhang:2005vt,Caldwell:2007cw,Guzik:2009cm,Bean:2010zq,Zhao:2010dz,Reyes:2010tr,Daniel:2010ky,Daniel:2010yt,Zhao:2012aw,Raveri:2014cka,Bellini:2015xja,Hojjati:2015ojt,Salvatelli:2016mgy,Joudaki:2016kym,Mueller:2016kpu,Zhao:2017cud,Amon:2017lia,Aghanim:2018eyx,DES:2018ufa,Noller:2018wyv,SpurioMancini:2019rxy,eBOSS:2020yzd,KiDS:2020ghu,DES:2021zdr} or even reconstruct the temporal behavior of certain models \cite{Raveri:2019mxg,Pogosian:2021mcs}, but canvassing the space of modified-gravity theories is challenging because that space is extremely large and difficult to constrain with currently available cosmological surveys.

In this paper we aim to answer a fundamental question: 
\begin{center}
    \textit{What happens when the data is analyzed assuming smooth dark energy and the universe is dominated by modified gravity? }
\end{center}
\noindent Such a scenario will clearly lead to an overall biased estimate of the inference of the cosmological model; see for example Figure 1 in Ref.~\cite{Huterer:2006mva}. Yet it would be very useful to know if modified-gravity theories lead to \textit{generic} shifts in the cosmological parameters relative to their true values. For example,  it could be that a departure of the equation of state $w$ relative to its \lcdm\ value of $-1$ indicates modified gravity. Or, that the currently observed Hubble tension --- the discrepancy between measurements of $H_0$ from the distance ladder and the CMB --- is a signature of modified gravity (something that a number of papers in the literature have explored, e.g. \cite{Lin:2018nxe, Braglia:2020auw}). It would be extremely useful to have knowledge of whether there are any \textit{generic} parameter shifts that modified gravity typically induces if analyzed assuming the standard unmodified model.

To address the highlighted question above, we opt for a forward-modeling approach. We wish to generate a large number of modified-gravity models, coming perhaps from different \textit{classes} of such models, and compute the cosmologically observable quantities. We then analyze those observables using some assumed future data, consisting of the cosmic microwave background, baryon acoustic oscillations, and type Ia supernova (these data are further discussed in Sec.~\ref{sec:data}). Crucially, when analyzing these data we assume unmodified-gravity, i.e. the \lcdm\ or the $w_0w_a$CDM model. We can thus assess the bias in all cosmological parameters, relative to their true values, due to the fact that data were analyzed using a wrong model. We then iterate the procedure many times. This informs us about 
what range of values for the standard (unmodified-gravity) cosmological parameters are inferred when the universe is subject to modified gravity.

One important decision in this procedure is to choose a general framework of modified-gravity theories from which to sample individual models.
Here we opt to utilize a familiar approach from particle physics (and, as of recently, cosmology) --- the Effective Field Theory (EFT). Here our approach is to utilize the EFT of Dark Energy (EFTDE) \cite{Park:2010cw,Gubitosi:2012hu,Bloomfield:2012ff}, where (universality) classes of models are established through a grouping of terms in the fundamental Lagrangian. This has the advantage that instead of considering one particular model at a time, one can consider an entire class of models with similar properties. One example of such a universality class in the EFTDE are the Horndeski models of modified gravity. In fact, here we will focus our investigation on the Horndeski sub-class of EFTDE models as described in Sec.~\ref{sec:EFTDE} below. 

Our procedure in this paper also includes a solution to a pesky technical problem: how to fit the eight-dimensional $w_0w_a$CDM models to each of the thousands of EFTDE models. This is computationally expensive because traditional Boltzmann-Einstein equation solvers used for this purpose such as \texttt{CAMB} are slow for what we are trying to do here. We thus employ and adapt an existing emulator package  to speed up this fitting process. This development enables us to obtain our numerical results with relatively modest computer resources. Most readily available cosmological emulators for the CMB power spectrum (such as \cite{Schneider2011} and \cite{SpurioMancini:2021ppk}) function for a fixed set of parameters -- usually the standard six cosmological parameters -- while our methodology of setting up the emulator allows a much greater freedom in including parameters.  

The paper is organized as follows. Sec.~\ref{sec:methodology} is divided into two parts and gives an overview of our overall methodology. The first half explains how we select a subset of Horndeski gravity models and compute cosmological observable quantities from them. The second half goes over methods (including a brief introduction on the emulation technique) used to reinterpret the data vectors generated by Horndeski models by fitting them with an unmodified-gravity $w_0 wa$CDM model. Sec.~\ref{sec:data} introduces the cosmological probes and assumed future experiment data used in the fitting process. Sec.~\ref{sec:results} discusses and summarizes the results. We conclude in Sec.~\ref{sec:concl}.



\begin{figure*}
\centering
\includegraphics[width=\textwidth]{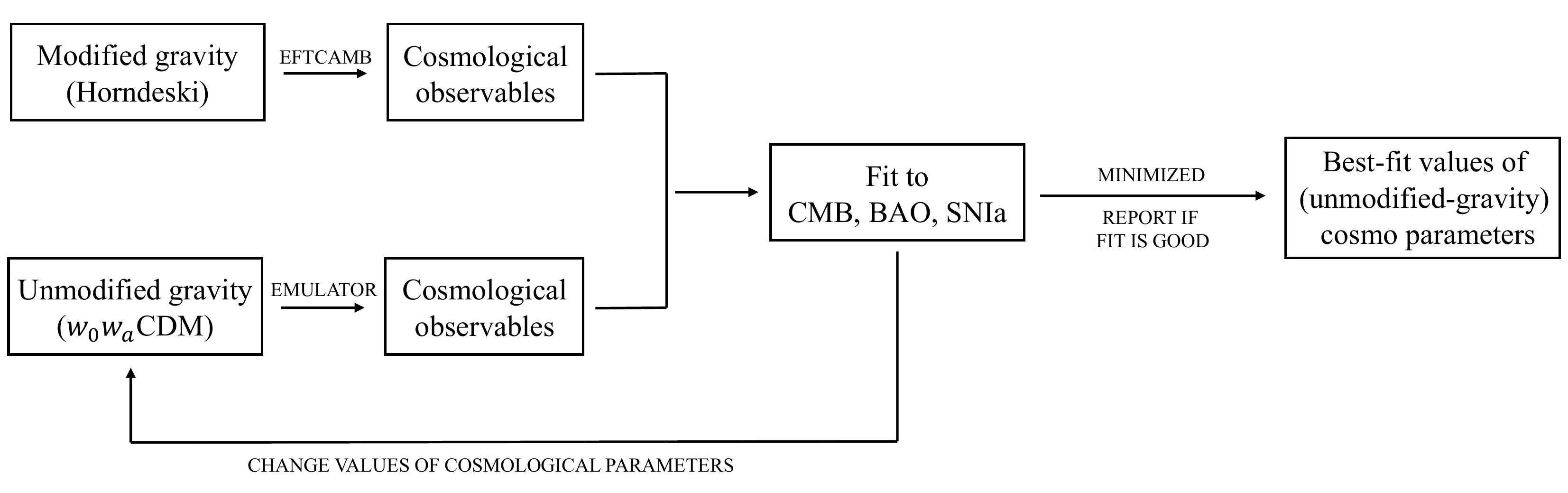}
\caption{A schematic describing our pipeline to  interpret and fit a modified gravity data vector with an (unmodified-gravity) dark energy model. We show the complete procedure for a single Horndeski data vector, corresponding to one point in our final best-fit parameter values in the plots that follow. We repeat this procedure procedure for thousands of Horndeski models.}
\label{fig:schematic}
\end{figure*}

\section{Methodology Overview} \label{sec:methodology}

As discussed in Sec.~\ref{sec:introduction}, we generate the data vector assuming a modified-gravity model, but analyze it assuming unmodified gravity in the $w_0w_a$CDM model.  Specifically, for each Horndeski data vector, we generate a CMB angular power spectrum predicted by this theory through a package \texttt{EFTCAMB}\footnote{\url{https://eftcamb.github.io}}, and also generate predictions for BAO and SNIa. Then, we fit to this synthetic data with $w_0w_a$CDM cosmological models. We record the best-fit parameters of such $w_0w_a$CDM model, and move on to the next iteration, selecting a new EFTDE model. Figure \ref{fig:schematic} shows our approach schematically. 

We now describe the key pieces of our approach: the modified-gravity theory to generate fake data, and the unmodified-gravity theory to analyze it with. For both modified and unmodified-gravity aspects of our analysis, we also discuss the numerical tools that enable the feasibility of our analysis.

\subsection{Generating data: modified gravity} \label{sec:EFTDE}

Inspired by the EFT formalism for Inflation by Cheung et. al. \cite{Cheung:2007st}, the EFTDE provides a universal description for all viable dark energy and modified gravity models \cite{Park:2010cw,Gubitosi:2012hu,Bloomfield:2012ff}
Working in unitary gauge, the EFTDE action takes the form \cite{Bloomfield:2012ff},
\begin{eqnarray}
\label{EFTDEaction}
    S &=& \int d^4x\sqrt{-g}\bigg[\frac{1}{2}m_0^2\Omega(t)R -\Lambda(t)-c(t)g^{00} + \frac{M_2^4(t)}{2}(\delta g^{00})^2\nonumber\\
    &-& \frac{\bar{M}_1^3(t)}{2}\delta K\delta g^{00}
    -\frac{\bar{M}_2^2(t)}{2}\delta K^2  -\frac{\bar{M}_3^2(t)}{2}\delta K_{\nu}^{\,\,\,\mu}\delta K_{\mu}^{\,\,\,\nu}\nonumber\\
    &+&\frac{\hat{M}^2(t)}{2}\delta R^{(3)}\delta g^{00}+m_2(t)\partial_ig^{00}\partial^ig^{00}
    +\mathcal{L}_m\bigg],
\end{eqnarray}
where $\delta g^{00} = g^{00} +1$ is the perturbation to the time component of the metric, $R^{(3)}$ is the perturbation to the spatial component, and $\delta K_{\mu\nu}$ is the perturbation of the extrinsic curvature.  The background evolution depends on three functions, $c(t)$, $\Lambda(t)$, and $\Omega(t)$. Two of the three can be constrained using the Einstein equations and are equivalent to the energy density and pressure. The third function, $\Omega(t)$, parameterizes the effect of modified gravity \cite{Bloomfield:2012ff}. In what follows we will take $\Omega =1$, thus explicitly fixing the background to $\Lambda$CDM\footnote{$m_0$ is the mass scale of the theory and is equivalent to $m_{\rm pl}$ when $\Omega(t) = 1$.}. The rest of the EFT functions describe perturbations about this background and correspond to observables that we are interested in when comparing to observations. For a summary of all models included in this very general formalism, refer to Table 1 in \cite{Linder:2015rcz}. Again, we note that the EFTDE includes such well-known simpler models as DGP and $f(R)$.

Here we specialize in a very broad subset of models captured by the EFTDE approach --- Horndeski models (for a general review of this class of models see \cite{Kobayashi:2019hrl} and references therein).  These models have been of particular interest because even if one does not take the EFTDE approach they have stable, second order equations of motion, leading to a well defined Cauchy problem and viable models of modified gravity. However, within the EFTDE approach, this is guaranteed from the outset. This universality class of models is obtained when the following relations are imposed on EFTDE functions
\begin{equation}
    2\hat{M}^2 = \bar{M}_2^2 = -\bar{M}_3^2; \qquad m_2 = 0.
    \label{eq:HordeskiGravity}
\end{equation}

We will be interested in the linear-theory predictions of Horndeski models as given by the EFTCAMB code  \cite{Hu:2014oga}. There, the EFTDE is described in terms of dimensionless parameters $\gamma_i$ defined as 
\begin{equation}
\begin{aligned}
\gamma_1&=\frac{M_2^4}{m_0^2H_0^2},\,\,\,\,\,\,\gamma_2=\frac{\bar{M}_1^3}{m_0^2H_0},\,\,\,\,\,\,\,\,\gamma_3=\frac{\bar{M}_2^2}{m_0^2},\\
\gamma_4&=\frac{\bar{M}_3^2}{m_0^2},\,\,\,\,\,\,\,\,\,\,\,\,\,\gamma_5=\frac{\hat{M}^2}{m_0^2},\,\,\,\,\,\,\,\,\,\,\,\,\,\,\gamma_6=\frac{m_2^2}{m_0^2}.
\end{aligned}
\label{eq: gammas}
\end{equation}
In terms of these new variables, the Horndeski models are obtained from the full EFTDE with these conditions
\begin{equation}
    2\gamma_5 = \gamma_3 = -\gamma_4; \,\,\, \gamma_6 = 0.
\end{equation}
Our approach is therefore to canvass through the possible Horndeski models by varying $\gamma_i(t)$ for $i=1, 2, 3$. 

There is an important caveat to our assumptions about the Horndeski parameter space. It has been argued that there exists a strong additional constraint on the parameter $\gamma_3$, based on the  comparison of the speed of light and gravitational-wave speed of propagation from the event GW170817 discovered by LIGO (see e.g. \cite{Kreisch:2017uet}). Because $\gamma_3$ is related to the speed of the gravitational wave $c_T$ (see e.g. \cite{Linder:2015rcz} and references therein), such a constraint would impose a strong prior that $\gamma_3$ is very close to zero. However, there are various theoretically motivated possible exceptions to this constraint \cite{deRham:2018red,Amendola:2018ltt,Battye:2018ssx}. With that in mind, and to make our analysis broadly applicable and not tied to specific theoretical models, we opt to keep $\gamma_3$ as a free parameter without any gravitational-wave-inspired prior. [To reinsert this prior, one could simply inspect and analyze our results evaluated for the small range of $\gamma_3$ around zero, although of course such an analysis will necessarily have a lower statistics than one where the $\gamma_3$ prior has been assumed from the beginning.]

In our approach, we require Horndeski models to successfully reproduce an approximate $\Lambda$CDM background and then focus on the connection between the perturbations and observations. That is, we set $w \simeq -1$, corresponding to $c \simeq 0$, $\Lambda \simeq$ constant and $\Omega \simeq 1$
in the EFTDE. This is a subset of Horndeski models, but corresponds to those consistent with a viable alternative to $\Lambda$CDM as required by data. Our approach is similar to that of the EFT of inflation where one assumes an inflationary background and then focuses on the perturbations (observables) \cite{Cheung:2007st}.

With the background constrained to a $\Lambda$CDM universe, we now consider allowed variations in the perturbations of our Horndeski models. Recall that there are three free time-dependent EFTDE functions in Horndeski gravity, $\gamma_i(t)$ for $i=1, 2, 3$. The first task is to parametrize the time-dependence of these functions, which we take as
\begin{equation}
    \gamma_i(a)=\gamma_{i,0}a,
\end{equation}
reproducing the CMB power spectra that are closest to current observations.  

Next, we determine the range of the coefficients $\gamma_{i,0}$. 
In Sec.~\ref{sec:analysis}, we describe how we set a $5\sigma$ requirement for each unmodified-gravity model as to be a good fit for the Horndeski model. By phenomenologically studying sample fits to various Horndeski models, we determine that the Horndeski parameter space restricted to the range
\begin{equation}
    \gamma_{1,0}\leq1;\,\,\,\,\, \gamma_{2,0}\leq0.1;\,\,\,\,\, \gamma_{3,0}\leq0.06,
    \label{eq:Horndeski-range}
\end{equation}
encompasses models that are sufficiently in correspondence to unmodified-gravity models, using criteria that we now describe.

\subsection{Analyzing data: unmodified gravity}\label{sec:analysis}

Our main goal is to fit simulated modified-gravity data using standard dark energy (unmodified-gravity) models. To be as general as possible, we fit $w_0w_a$CDM cosmological models to the data, with parameters
\begin{equation} \label{eq:params}
    \{p_i\}=\left\lbrace\omega_{b}, \omega_c, H_{0}, \ln (10^{10}A_s), n_{s}, \tau_{\rm reio}, w_0, w_a\right\rbrace \,,
\end{equation}
where $\omega_b\equiv\Omega_b h^{2}$ is the physical baryon density, $\omega_c \equiv \Omega_c h^2$ is the physical cold dark matter density, $H_0$ is the Hubble constant, $A_{s}$ is the amplitude of the primordial power spectrum at pivot wave number $k_{\rm piv}=0.05\ {\rm Mpc}^{-1}$, $n_s$ is the scalar spectral index, $\tau_{\rm reio}$ is the optical depth to reionization, and ($w_0, w_a$) are the parameters describing the dark energy equation of state.

For each Horndeski data vector generated using \texttt{EFTCAMB} with assumptions as described in Sec.~\ref{sec:EFTDE}, we need to find the best-fit $w_0w_a$CDM model. We thus need to be able to produce the supernova and BAO observables (distances and the Hubble parameters) and the CMB angular power spectrum in $w_0w_a$CDM models many times for a single Horndeski model. Calculating distances is straightforward, while the CMB temperature and polarization angular power spectra are typically obtained using the standard Boltzmann-Einstein solver \texttt{CAMB}. Here we employ an emulator due to computational cost reasons explained above.

Given a single Horndeski data vector and predictions from unmodified-gravity models, we minimize the total $\chi^2$, defined as a sum of chi-squareds for each cosmological probe in Sec.~\ref{sec:data} and thus find the best-fitting parameters. To carry out chi-squared minimization in our eight-dimensional parameter space given in Eq.~(\ref{eq:params}), we adopt  \textsf{iminuit}\footnote{\url{https://iminuit.readthedocs.io/en/stable/}}. This optimizer allows us to restart the minimization process from the ending point of the last minimization, re-doing the minimization five times for each EFTDE model to improve the result. The allowed ranges for each parameter to explore is set to be 5\% smaller than the parameter range specified in Table \ref{tab:parameters}. 

As alluded to in Sec.~\ref{sec:EFTDE}, we wish to only use reasonably good fits to our Horndeski data vectors, as an analysis resulting in a bad fit to the data would simply not be allowed to proceed in a realistic situation. To that end, we only accept best-fit $w_0w_a$CDM models that have a minimized $\chi^2$ within 5$\sigma$ of the expectation for a chi-square distribution of $N_{\rm dof}$ degrees of freedom. Our simulated cosmological data, described below in Sec.~\ref{sec:data}, have $N_{\rm dof}=7492$\footnote{We used $3 \times 2500$ multipoles from temperature and polarization spectra respectively as our data, and it was constrained by 8 parameters as listed in Table~\ref{tab:parameters}.}. Recall that our simulated Horndeski data vectors are noiseless, so that a perfect fit would have $\chi^2=0$. With this information, the "5-$\sigma$" limit to a cosmological fit corresponds to chi-square limit of
\begin{equation}
    \chi^2<650 \quad\mbox{(acceptable fit)}.
    \label{eq:chisq_crit}
\end{equation}

If the best fit to a given Horndeski model is worse than this, we judge that such a model would not be interpreted as a viable cosmological model. We also exclude results for models where one or more parameters reach the upper or lower bounds of their respective parameter range given in Table~\ref{tab:parameters} as it indicates that this model cannot be fitted by a $w_0 w_a$CDM model within the range of current measurements well; this affects about 21 percent of Horndeski models that we considered. 

In our model-fitting  procedure, the main challenge is the significant computational cost. Consider that  \texttt{CAMB}\footnote{\url{https://camb.info}} takes about 1.5 second to produce a $w_0w_a$CDM CMB angular power spectrum. For a single Horndeski model, the minimizer requires of order 1,000 $w_0w_a$CDM model evaluations, and our overall goal is to produce results for 10,000 or more Horndeski models. To addrress this challenge we constructed an emulator to generate model predictions for $w_0 w_a$CDM cosmologies. 
An emulator is essentially an interpolator. Given a set of grid points in an $N$-dimensional parameter space and corresponding outcomes evaluated at these points, the emulator interpolates to produce an expected outcome on arbitrary points off the grid (but still within its boundaries). In our case, the grid is the eight-dimensional parameter space listed in Eq.~(\ref{eq:params}). Since the spectrum is obtained through interpolation, and not from solving the Boltzmann-Einstein equation, this method generates spectra much faster. The emulator we developed builds on the  \texttt{EGG}\footnote{\url{https://github.com/lanl/EGG}} package. 

\begin{table}[H]
    \caption{Fiducial values of cosmological parameters and their ranges used in training the emulator}
  \begin{center}
    \begin{tabular}{c|c|c}
      
      \hline 
      \textbf{Parameter} & \textbf{Fiducial value} & \textbf{Parameter range} \\

      \hline
      $\Omega_b h^2$ & 0.02222 & (0.02147, 0.02297)  \\
      $\Omega_c h^2$ & 0.1197 & (0.1137, 0.1257)  \\
      $A_s$ & 2.196 $\times 10^{-9}$ & (1.132 $\times 10^{-9}$, 2.703 $\times 10^{-9}$ ) \\
      $H_0$ & 67.5 & (64.8, 70.2)  \\
      $n_s$ & 0.9655 & (0.9445, 0.9865)  \\
      $\tau$ & 0.06 & (0.0235, 0.0965)  \\
      \hline
      $w_0$ & -1 & (-1.5, -0.5)  \\
      $w_a$ & 0 & (-0.5, 0.5)  \\
      \hline 
    \end{tabular}
  \end{center}
  \label{tab:parameters}
\end{table}

\begin{itemize}
    \item \textit{Parameter ranges:} 
The prior range for each of the first six parameters in Eq.~(\ref{eq:params}) is set to $\pm 5 \sigma$ around their fiducial values, where $\sigma$ is the 68\% marginalized error on each corresponding parameter from the Planck 2018 analysis using the \texttt{Plik} likelihood \cite{Planck:2018vyg}. For the two dark energy parameters $w_0$ and $w_a$, we adopt ranges $-1.5 \leq w_0 \leq -0.5$ and $-0.5 \leq w_a \leq 0.5$. A summary of all parameter ranges are in Table  \ref{tab:parameters}.
\item \textit{Parameter grid values:} A uniform grid is not ideal as, for a reasonable number of values in each parameter, it leads to a large number of grid points and slow emulator training. Therefore, we employ the Latin Hypercube sampling (LHS) which is known to be very efficient for emulators \cite{Heitmann2009}.  The points in LHS are stratified along the direction of each axis in a multi-dimensional space. This design is mathematically equivalent to forming a $n\times m$ matrix such that every column of this matrix is a unique permutation of $\{1,...,n\}$. There are a number of strategies to design an LHS\footnote{We did not opt for the commonly used orthogonal-array Latin hypercube (OALH) design. This is because using OALH, one relies on the existing library of orthogonal arrays, and the latter does not offer much flexibility to change the number of parameters and the number of samples (i.e. grid points). Specifically, there exist only a few available orthogonal arrays for an eight-dimensional parameter space, and the allowed sample numbers for these arrays are too low for our purposes. The strategy we adopt, as discussed in the text, is not as optimal as the OALH design in its coverage of the parameter space, but its performance can be easily improved through increasing the number of grid points.}, and the one we use is provided by a python package \texttt{pyDOE}\footnote{\texttt{Designs of Experiments for Python}, \url{https://pythonhosted.org/pyDOE/randomized.html\#latin-hypercube}}. This package allows us to specify the number of parameters and the number of grid points with much greater flexibility. 
\item \textit{Training:} To ``train'' an emulator is to assign the corresponding outcomes to the grid points. Here, we use \texttt{CAMB} to calculate the CMB temperature and polarization angular power spectra (TT, EE, and TE) and assign them to the corresponding grid points. During training, the emulator uses a Markov chain Monte Carlo type process to find and optimize an interpolative function that describes the nonlinear relationship between the grid points and their corresponding CMB power spectra.

\item \textit{Testing emulator's performance:} The performance of an interpolation under a given LHS setup can be determined quantitatively by comparing the interpolated power spectrum at an arbitrary point in parameter space with the one generated directly by \texttt{CAMB}. Adopting a test similar to the one used in \cite{Schneider2011}, we randomly selected 100 points from the allowed parameter space in Table~\ref{tab:parameters} and calculated the fractional difference between the angular power spectrum interpolated by the emulator and the power spectrum generated by \texttt{CAMB}. For the temperature power spectrum, the emulator's fractional errors within the first and third quartile are 0.3\% for multipoles $\ell > 8$. For the polarization power spectra EE and TE, the fractional errors are 0.5\% for $\ell > 25$ and 3.5\% for $\ell > 55$ respectively. 

The performance of the interpolation is mostly determined by the number of grid points in the LHS design and the number of MCMC iterations when training the emulator. A larger number of grid points and a higher number of steps in the MCMC-type process during training would both improve the performance of interpolation, but at the cost of a slower evaluation per model. In this work, we use 570 grid points and 1000 iterations. With the current setup, each interpolation takes about 0.3 seconds to finish, which is five times faster than using CAMB.

\end{itemize}

\section{Simulated Data} 
\label{sec:data}

In this section, we will discuss the probes and experiment specifics we used to determine the best-fit values of dark energy parameters $w_0$ and $w_a$. 

We use cosmic microwave background, baryon acoustic oscillations (BAO), and type Ia supernovae (SN Ia) as our data. In this first paper on the topic, we opt not to use weak gravitational lensing or galaxy clustering. As mentioned in the introduction, this is due to the significant additional complexity in modeling clustering, which for starters one typically needs to restrict to linear scales only in modified-gravity models as obtaining reliable nonlinear predictions is  very challenging.  It is our goal to set up a robust proof-of-principle analysis pipeline with the CMB, BAO and SN Ia alone. In a future publication, we will add the galaxy clustering and weak lensing (and, ideally, the full "3x2" pipeline that also includes galaxy-galaxy lensing).

A summary of the probes used can be seen in Table \ref{table:probes}. We now describe them in more detail. 

\begin{table*}
    \caption{A summary table of the probes and data sets used to determine the best-fit parameters for a certain EFTDE model.}
  \begin{center}
    \begin{tabular}{c|c|c|c|c}
      
      \hline 
      \textbf{Probes} & \textbf{Experiment} & \textbf{Measurements} & \textbf{Details} & \textbf{Data error} \\
      \hline
      \multirow{3}{*}{CMB} & \multirow{3}{*}{Stage-4} &  \multirow{3}{*}{angular power spectrum $C_\ell$} & from $l = 2$ to $l = 2500$ & \multirow{3}{*}{Eq.~\ref{cov}} \\
      & & & only had the temperature (TT) components before \\
      & & & now added polarization (EE and TE)\\
      \hline
      \multirow{3}{*}{SNIa} & \multirow{2}{*}{WFIRST} & \multirow{2}{*}{apparent magnitude $m(z)$} & 16 effective supernovae in redshift bins of size $0.1$ & \multirow{2}{*}{Eq.~\ref{tot}} \\
      & & & from $z = 0.1$ to $z = 1.6$ with ~0.4\% error \\\cline{2-5}
      & Pan-STARRS1 & apparent magnitude $m(z)$ & 870 supernovae from $z=0.00508$ to $z=1.06039$ & Ref.~\cite{Scolnic:2017caz} \\
      \hline
      \multirow{2}{*}{BAO} & \multirow{2}{*}{DESI} & angular diameter distance $D_A(z)$  & 13 redshift bins of size $0.1$ & \multirow{2}{*}{Ref.~\cite{DESI:2016fyo}} \\
      & & Hubble parameter $H(z)$ & from $z = 0.65$ to $z = 1.85$ \\
      \hline 
    \end{tabular}
  \end{center}
  \label{table:probes}
\end{table*}

\subsection{CMB}

We assume a CMB survey modeled on expectations from CMB-S4 \cite{Li:2018epc}. The survey covering 40\% of the sky, with other specifications given below. We utilize scales out to maximum multipole $\ell_{\rm max}=2500$, consistent with the cutoff in Planck 2018 results\cite{Planck:2018vyg}. Assuming a Gaussian likelihood $\mathcal{L}$, the chi squared, $\chi^2\equiv -2\ln\mathcal{L}$, is given by 

\begin{equation}
    \chi^2_{\rm CMB} = \sum^{\ell = 2500}_{\ell = 2} \left(\textbf{C}_\ell^{\rm data} - \textbf{C}_\ell^{\rm th}\right)^{T} \text{ Cov}_\ell^{-1} \left(\textbf{C}_\ell^{\rm data} - \textbf{C}_\ell^{\rm th}\right),
\end{equation}
where $\textbf{C}_\ell^{\rm th}$ is the data-vector corresponding to theory ($w_0w_a$CDM) prediction,  and $\textbf{C}_\ell^{\rm data}$ are the data which, recall, are produced assuming the EFT model.  Both the theory and the data $\textbf{C}_\ell$ are composed of parts corresponding to temperature-temperature (TT), temperature-polarization (TE), and polarization-polarization (EE) correlations:
\begin{equation}
    \textbf{C}_\ell\equiv
    \left(\begin{matrix} 
    C_\ell^{TT} \\[0.2cm] 
    C_\ell^{EE} \\[0.2cm] 
    C_\ell^{TE} 
    \end{matrix} \right).
\end{equation}
The overall covariance matrix $\mathbf{\rm Cov}_\ell$ is diagonal between the different multipoles. At each multipole, the covariance for the data vector $\textbf{C}_\ell^{\rm data}$ is given by (e.g.\ \cite{Li:2018epc})

\begin{eqnarray}
    \mathbf{\rm Cov}_\ell & =& \frac{2}{(2\ell + 1) f_{\rm sky}}\\[0.2cm]
    &\times&
    \left( \begin{matrix} (\Tilde{C}_\ell^{TT})^2 & (\Tilde{C}_\ell^{TE})^2 & \Tilde{C}_\ell^{TT} \Tilde{C}_\ell^{TE}\\[0.2cm] (\Tilde{C}_\ell^{TE})^2 & (\Tilde{C}_\ell^{EE})^2 & \Tilde{C}_\ell^{EE} \Tilde{C}_\ell^{TE}\\[0.2cm] \Tilde{C}_\ell^{TT} \Tilde{C}_\ell^{TE} & \Tilde{C}_\ell^{EE} \Tilde{C}_\ell^{TE} & \frac{1}{2}[ (\Tilde{C}_\ell^{TE})^2 + \Tilde{C}_\ell^{TT} \Tilde{C}_\ell^{EE}] \end{matrix} \right).
    \nonumber
    \label{cov}
\end{eqnarray}
\noindent
The elements of this covariance matrix are explicitly
\begin{equation}
    \begin{aligned}
    \Tilde{C}_\ell^{TT} & = C_\ell^{TT} + N_\ell^{TT}  \\
    \Tilde{C}_\ell^{EE} & = C_\ell^{EE} + N_\ell^{EE}  \\
    \Tilde{C}_\ell^{TE} & = C_\ell^{TE},
    \end{aligned}
\end{equation}

\noindent
and the noise terms are
\begin{equation}
    \begin{aligned}
    N_\ell^{TT} & =  \Delta^2_T \exp \left[ \frac{\ell(\ell+1) \theta^2_{\rm FWHM}}{8 \ln{2}} \right] \\
    N_\ell^{EE} & =  2 \times N_\ell^{TT},
    \end{aligned}
\end{equation}

\noindent
where $\Delta_T = 1\,\mu K$, $\theta_{\rm FWHM} = 8.7 \times 10^{-4}$ radians, and assume $f_{\rm sky} = 0.4$, using the specifics of the Stage-4 experiment \cite{Li:2018epc}.

We generate the data vector $\textbf{C}_\ell^{\rm data}$ (for each $\ell$) using \texttt{EFTCAMB}, for a given cosmological model as discussed in Sec.~\ref{sec:EFTDE}. This is an important step, as CMB is the only part of our simulated data that is directly affected by modified gravity.

We generate \textit{noiseless} data vectors --- that is, the final $\textbf{C}_\ell^{\rm data}$ used in the likelihood are precisely centered on theory, with no stochastic noise. This assumption is justified because we are not interested in statistical errors on the infered parameters, but rather only at the best-fit parameters (for a given simulated Horndeski model). Had we included stochastic noise, we could have still obtained the results that we are after, but it would have required running a number of statistical realizations of data vectors for a given Horndeski model in order to account for stochasticity in the data. 

\subsection{SNIa}

Type Ia supernovae (SNIa) are sensitive to distances alone. Because in our generated data we fix the background cosmology to \lcdm\ and only vary the perturbations according to modified gravity, SNIa data vector is not directly sensitive to modified gravity. Nevertheless, SNIa are very useful in pinning down the cosmological parameters and breaking degeneracies between them, and thus helping isolate the effects of modified gravity on data analyzed assuming $w_0w_a$CDM.

Assuming again a gaussian likelihood, the chi squared for SNIa measurements is determined by
\begin{eqnarray}
   \chi^2_{\rm SN} (\{p_i\}, \mathcal{M}) & = & (\textbf{m}^{\rm data} - \textbf{m}^{\rm th})^T \mathbf{\rm Cov}^{-1} (\textbf{m}^{\rm data} - \textbf{m}^{\rm th}),
    \nonumber
\end{eqnarray}
where $\textbf{m}^{\rm data}$ is the apparent magnitude of simulated data which is calculated based on the cosmology in each fit to the Horndeski model. The theoretical magnitude $\textbf{m}^{\rm th}$ is, conversely, calculated based on the fiducial $w_0w_a$CDM  cosmological model:
\begin{equation}
    m^{\rm th} (z) = 5 \text{log}_{10} [H_0 d_L (z, \{p_i\})] + \mathcal{M}
    \label{m_th}
\end{equation}
where $d_L$ is the luminosity distance, and $\mathcal{M} = M - 5 \text{log}_{10} (H_0 \times 1 \text{Mpc}) + 25$ is a nuisance parameter that always needs to be marginalized over in a SNIa analysis. We can analytically marginalize over $\mathcal{M}$ and obtain a marginalized effective $\chi^2$
\begin{equation}
    \chi^2_{\rm SN,\, marg} = a - \frac{b^2}{c},
\end{equation}
where 
\begin{equation}
    \begin{aligned}
    a & =  (\textbf{m} - \textbf{m}^{th})^T \mathbf{\rm Cov}^{-1} (\textbf{m} - \textbf{m}^{th}) \\[0.2cm]
    b & =  \textbf{1}^T \mathbf{\rm Cov}^{-1} (\textbf{m} - \textbf{m}^{th})  \\[0.2cm]
    c & =  \textbf{1}^T \mathbf{\rm Cov}^{-1} \textbf{1}, 
    \end{aligned}
\end{equation}
where $\textbf{1}$ is a unit vector.

We employed the SNIa redshift bins and the covariance matrix as forecasted for the WFIRST satellite \cite{Hounsell:2017ejq}. The covariance matrix is diagonal between different bins, and is calculated as a combination of systematic and statistical errors. In a given redshift bin,
\begin{equation}
    \sigma_{\rm tot} = (\sigma_{\rm sys}^2 + \sigma_{\rm stat}^2)^{1/2}, 
    \label{tot}
\end{equation}
where
\begin{equation}
\begin{aligned}
    \sigma_{\rm sys} & =  \frac{0.01 (1+z)}{1.8} \label{sys}  \\[0.2cm]
    \sigma_{\rm stat} & =  \frac{(\sigma_{\rm meas}^2 + \sigma_{\rm int}^2 + \sigma_{\rm lens}^2)^{1/2}}{N_{\rm SN}^{1/2}}.
\end{aligned}
\end{equation}
Here, $\sigma_{\rm meas} = 0.08$, $\sigma_{\rm int} = 0.09$, $\sigma_{\rm lens} = 0.07z$, and $N_{\rm SN}$ is the number of supernovae in that redshift bin.

We have also incorporated redshift bins and the corresponding covariance matrix from measurements at low redshift by Pantheon dataset \cite{Scolnic:2017caz}, which includes 870 supernovae. The covariance matrix for this data set is diagonal, and the error at each redshift is given by Pantheon as well.

\subsection{BAO}

Baryon acoustic oscillations (BAO) --- wiggles in the matter power spectrum due to photon-baryon oscillations prior to recombination --- are a powerful cosmological probe. Much like SNIa, they probe geometry, and are sensitive to the angular-diameter distance $D(z)$ and Hubble parameter $H(z)$ evaluated at the redshift of tracer galaxies in question. Often, the general analysis of the BAO provides precisely these "compressed quantities" for one or more effective redshifts, which in turn can be used to constrain a cosmological model. 

Here we assume the $D(z)$ and $H(z)$ measurements that are forecasted to be measured DESI experiment \cite{DESI:2016fyo}.
The measurements of both the distances and the Hubble parameters are each reported separately in 13 redshift bins; we thus organize these measurements in  data vectors $\textbf{D}$ and $\textbf{H}$ that each have 13 elements. As before, we generate synthetic noiseless data ($\textbf{D}^{\rm data}$ and $\textbf{H}^{\rm data}$) assuming Horndeski models, and analyze it using theoretically computed quantities ($\textbf{D}^{\rm th}$ and $\textbf{H}^{\rm th}$) that assume the $w_0w_a$CDM model. 

The goodness-of-fit for BAO is written down in a similar way as for the CMB and SNIa 
\begin{equation}
    \begin{aligned}
    \!\!\!\!\chi^2_{\rm BAO}(\{p_i\}) & = (\textbf{D}^{\rm data} - \textbf{D}^{\rm th})^T \textbf{\rm Cov}_D^{-1} (\textbf{D}^{\rm data}  - \textbf{D}^{\rm th}) \\[0.1cm]
     & + (\textbf{H}^{\rm data} - \textbf{H}^{\rm th} )^T \textbf{\rm Cov}_H^{-1} (\textbf{H}^{\rm data} - \textbf{H}^{\rm th} ),
    \end{aligned}
\end{equation}
where $\textbf{\rm Cov}_D$ and $\textbf{\rm Cov}_D$ are respectively the $13\times 13$ covariance matrices for the distance and Hubble parameter measurements, which are diagonal. We adopt these matrices also from DESI forecasts \cite{DESI:2016fyo}.



\section{Results and Discussions} \label{sec:results}

\begin{figure*}
    \centering
    \includegraphics[width=15cm]{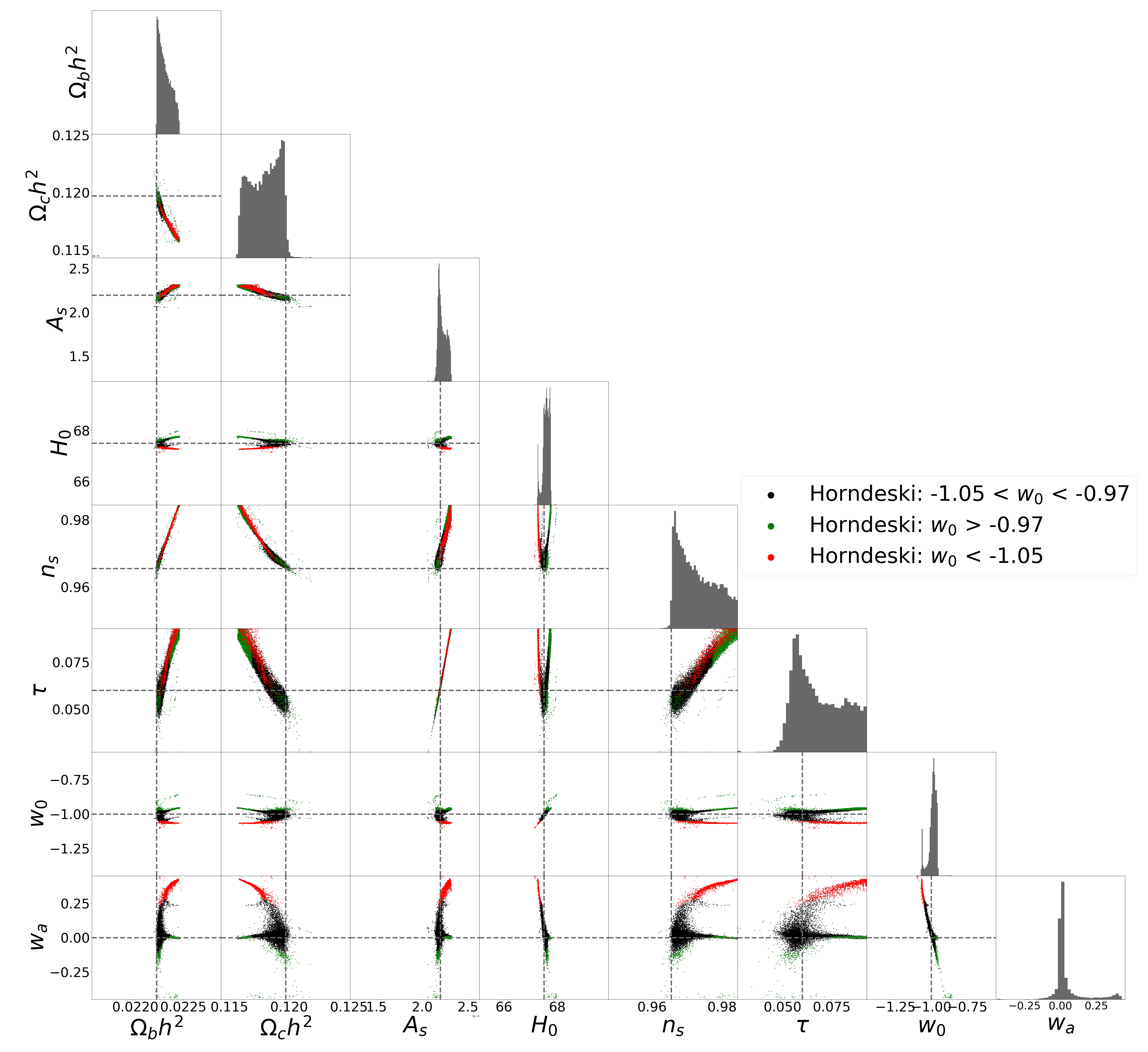}
    \caption{Best-fit values and histograms of cosmological parameters and dark energy sector parameters obtained from fitting to 15186 Horndeski models with a $w_0 w_a$CDM cosmology. Branches shown in the panels along the rows of $H_0$, $w_0$ and $w_a$ can be separated by values of $w_0$, as indicated by red and green points.}
    \label{fig:triangular}
\end{figure*}

Our results are summarized in Fig.~\ref{fig:triangular}. Here we show the eight-dimensional space of $w_0w_a$CDM models that were fit to Horndeski data vectors.  Each point corresponds to values of the best-fit $w_0w_a$CDM model for a given Horndeski model. We show results for a total of 15186 Horndeski data vectors which passed our criteria laid out in Sec.~\ref{sec:analysis}. We show all possible 2D planes of cosmological parameters, as well as histograms of the distributions in each parameter on the diagonal. The axis limits are chosen so that they indicate the range within which each parameter is allowed to vary during the minimization. The grey crosshair in each panel indicates our fiducial cosmology (see Table \ref{tab:parameters}), which corresponds to the background cosmology we set in all our Horndeski models. 

Note specifically that Fig.~\ref{fig:triangular} does \textit{not} show any kind of parameter constraint --- that is, no "error bars" are represented here. Rather, in each parameter panel of the Figure, the distribution of points relative to the crosshair demonstrates how values of the respective parameters shift relative to their true values when modified gravity (Horndeski) theories are incorrectly interpreted as dark energy ($w_0w_a$CDM).  Recall also that these fits are only performed for $w_0w_a$CDM  models that are decent fits to Horndeski data vectors, judged by the criterion in Eq.~(\ref{eq:chisq_crit}), mimicking the decision point that would be applied in an analysis of real data. Finally, the density of points in Fig.~\ref{fig:triangular} is not particularly important, as it merely reflects the metric on our prior in the space of models (e.g.\ the fact that we used a flat prior in the parameters $\gamma_i$ rather than, say, a log prior). What we are interested instead is the overall extents and shapes of the clouds of points.

The most apparent observation from Fig.~\ref{fig:triangular} is that the biases in $w_0w_a$CDM parameters, relative to their true values, carve out very specific directions in the parameter space. Table~\ref{table:parameter shifts} summarizes the directions in which the parameters are shifted. The specific shifts are generally unsurprising, as we would guess that there exist specific degeneracies between Horndeski models and $w_0w_a$CDM parameters where the former can be interpreted as the latter. Nevertheless, the precision to which the $w_0w_a$CDM biases are carved out in their respective parameter spaces is remarkable.

\begin{figure*}[t]
    \centering
    {\includegraphics[width=9.5cm]{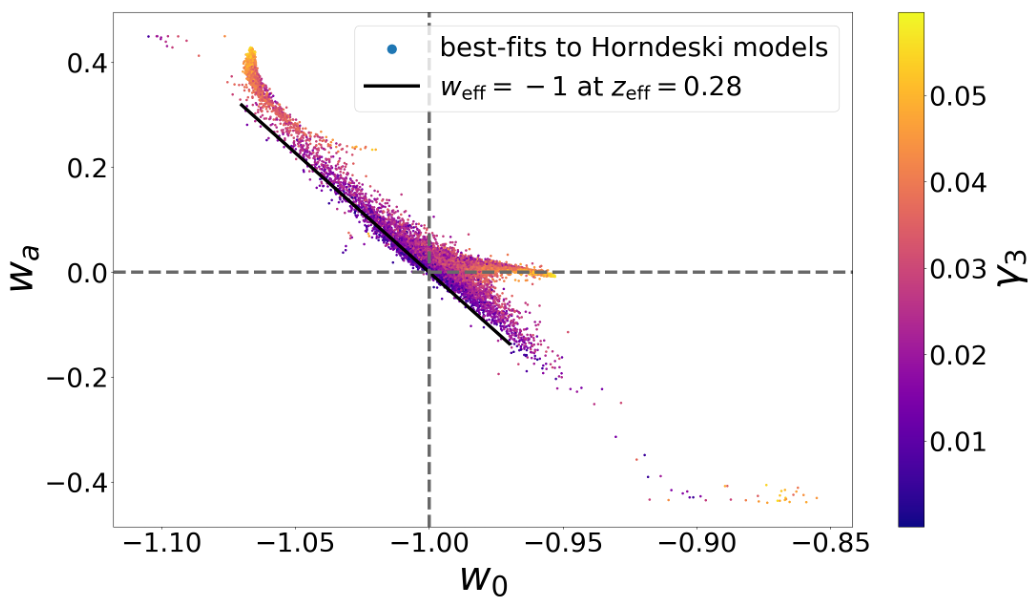}}
    {\includegraphics[width=8cm]{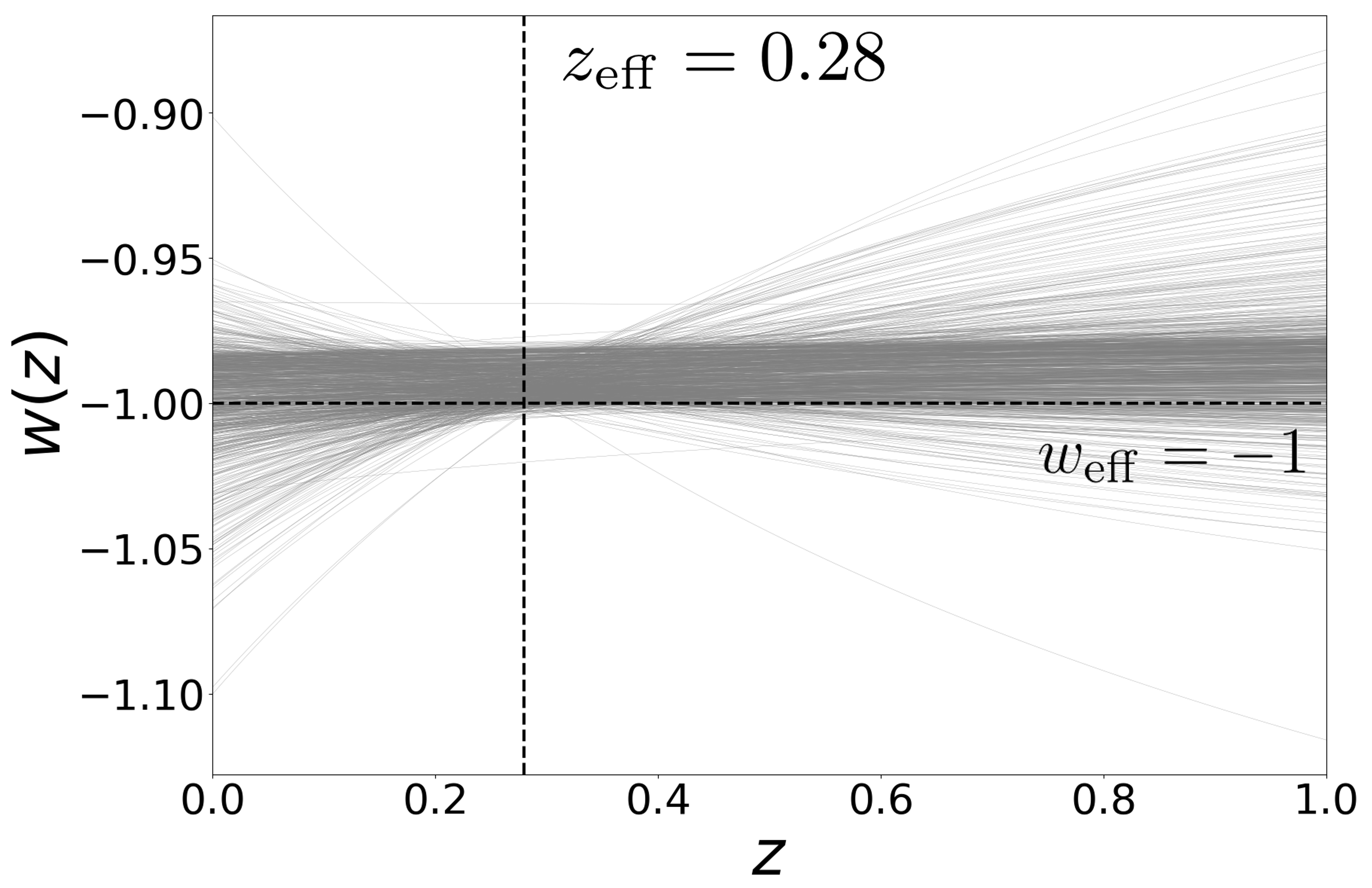}}
    \caption{\textit{Left panel:} The $w_0$-$w_a$ plane from Figure \ref{fig:triangular}, where each point is colored by the $\gamma_3$ function of the corresponding Horndeski data vector that was fitted with a $w_0 w_a$CDM cosmology. \textit{Right panel:} Equation of state $w(z)$ for 1000 randomly selected models (corresponding to a subset of points in the purple-pink region in the left panel). Notice that the equation-of-state curves intersect around an effective redshift $z_{\rm eff} = 0.28$, at the value of the effective equation of state typically slightly larger than $w_{\rm eff} = -1$.}
    \label{fig:w0wa}
  \end{figure*}

The next most noticeable feature of our results are the branchings in the $w_0w_a$CDM parameter biases. In other words, biases in the parameters trace out multiple (two or three) directions in several 2D parameter planes.  This indicates multiple degeneracy directions between shifts in the $w_0w_a$CDM space and Horndeski models. A very general quantitative expectation for this multi-modality is difficult to establish, but we have nevertheless  explored this in some detail. We found that the value of the parameter $w_0$ --- dark energy equation of state value today ---  is a good predictor for the branchings. Specifically, we found that modified-gravity models that are best fit with, respectively, $w_0<-1.05$ and $w_0>0.97$, lead to two prominent branches that are evident in a number of 2D planes, and that are labeled with green and red points respectively in Fig.~\ref{fig:triangular}. Conversely, models fit with $-1.05<w_0<-0.975$, labeled with black points, form the "core" of the distribution, at the nexus of the two branches. 

Closing the analysis of Fig.~\ref{fig:triangular}, note that the overall biases in the standard-model parameters are, very roughly, comparable to the current statistical uncertainties in these parameters. For example, the range of the scalar spectral index, roughly $[0.96, 0.98]$, is somewhat larger than its present statistical uncertainty, while that in the Hubble constant, $[66.86, 68.43]$, is also somewhat larger than the constraints from Planck 2018 analysis \cite{Planck:2018vyg}. This is not particularly surprising as we have only shown models whose fit to Horndeski data vectors is "good" as quantified in terms of near-future experimental errors. Nevertheless, this tells us that future constraints on these parameters will likely favor a subset of models shown in Figure \ref{fig:triangular}. Future data may thus indicate whether a specific sub-class of modified-gravity models lurks in the data.

Of particular interest to cosmologists is the measured value of the equation-of-state parameters $(w_0, w_a)$. Can these measured values indicate the presence of modified gravity? To help answer this question, we enlarge and display Fig.~\ref{fig:triangular}'s $w_0-w_a$ plane in the left panel Fig.~\ref{fig:w0wa}. First, note that the $w_0$ and $w_a$ values of best-fit unmodified-gravity models are mutually highly correlated. This is entirely expected, as the physically relevant quantity is $w(z)$ at the redshift where best constrained by the data --- the effective, or "pivot" redshift \cite{Huterer:2002hy,Linder:2006xb}. In fact, it turns out that our range of Horndeski models given by Eq.~(\ref{eq:Horndeski-range}), the largely one-dimensional direction of best-fit models in $w_0-w_a$ plane is
\begin{equation}
    w_{\rm eff} = w_0 + w_a (1-a_{\rm eff})\simeq -1
\label{eq:weff}
\end{equation}
with the effective scale factor $a_{\rm eff}=0.78$ or redshift $z_{\rm eff}=0.28$. Therefore, the best-fit models do allow variation in $w_0$ and $w_a$, but constrained such that it falls along the direction pointed by a constant $w$ at the effective redshift. To illustrate this, the black line in the left panel of Fig.~\ref{fig:w0wa} follows combinations of $w_0$ and $w_a$ that give $w_{\rm eff} = -1$ and $z_{\rm eff} = 0.28$ based on Eq.~(\ref{eq:weff}). Note that most best-fit models are actually slightly above the black dashed line, indicating that $w_{\rm eff}$ is slightly larger than $-1$. 
The linear relation in Eq.~(\ref{eq:weff}), along with the effective scale factor $a_{\rm eff}$ of approximately the size that we find, are respectively expected and predicted in the present scenario when the equation-of-state parameters are working to match the distance to the surface of last scattering imposed by CMB data \cite{Linder:2007ka}.

We shed more light on what best-fit $(w_0, w_a)$ values are favored as fits to Horndeski models in the right panel of Fig.~\ref{fig:w0wa}. Here, each curve represents the function $w(z)$ (in the $w_0, w_a$ model) for each corresponding (purple or pink-colored) point in Fig.~\ref{fig:w0wa}. Notably, most best-fit $w(z)$ curves intersect around the effective redshift $z_{\rm eff}=0.28$, the value that is indicated with a vertical black dashed line.  
  


It is also instructive to look at the overall extent of the distribution of models in the left panel of Fig.~\ref{fig:w0wa}. The coverage of the $w_0-w_a$ "island" is highly non-uniform, with more models with a positive $w_a$ than negative. We obtain additional information by plotting the $\gamma_3$ parameter from Eq.~\ref{eq: gammas} for each model, which dominates how far that Horndeski data vector's departure from our background $\Lambda$CDM cosmology is. As expected, lower values of $\gamma_3$ (i.e. models that resemble the $\Lambda$CDM background most) forms the core of the distribution, while models with higher values of $\gamma_3$ have larger deviations in $(w_0, w_a)$ and tend to either aggregate in the branch favoring a higher value of $w_0$ and $w_a$ around zero, or at the upper left tip which favors the lowest values of $w_0$ but the highest ones of $w_a$.


The left panel of Fig.~\ref{fig:w0wa} also shows a branching in the distribution of models in the $w_0-w_a$ plane, though weaker than the more prominent ones in the full 8D parameter space seen in Fig.~\ref{fig:triangular}. We did not pursue understanding this feature, given hat it is not extended, and probably encodes subtle correlations between dark energy parameters $(w_0, w_a$) and Horndeski model parameters when the former are enforced to fit the latter.


\begin{table}[t]
    \caption{Summary of the trends in the inferred cosmological parameters when modified-gravity (Horndeski) models are interpreted within the context of unmodified gravity --- either in $w_0 w_a$CDM or $\Lambda$CDM cosmology. For each parameter, we show the percentage of best-fit values larger/smaller than the true (input) value. Parameters whose best-fit values are  overwhelmingly shifted in the same direction are highlighted in red.}
  \begin{center}
    \begin{tabular}{c|c|c|c|c}
      
      \hline 
      & \multicolumn{2}{c}{$\boldsymbol{w_0 w_a}$\textbf{CDM}} & \multicolumn{2}{|c}{$\boldsymbol{\Lambda}$\textbf{CDM}} \\

      \hline
      Compared to & \multirow{2}{*}{\% Larger} & \multirow{2}{*}{\% Smaller} & \multirow{2}{*}{\% Larger} & \multirow{2}{*}{\% Smaller} \\
      fiducial value & & & \\
      \hline
      $\Omega_b h^2$ & \cellcolor{redD}99.7 & 0.3 & \cellcolor{redD}99.9 & 0.1 \\
      $\Omega_c h^2$ & 2.8 & \cellcolor{redD}97.2 & 1.1 & \cellcolor{redD}98.9 \\
      $A_s$ & 62.3 & 37.7 & 35.2 & 64.8 \\
      $H_0$ & 78.6 & 21.4 & \cellcolor{redD}99.2 & 0.8 \\
      $n_s$ & \cellcolor{redD}99.2 & 0.8 & \cellcolor{redD}99.97 & 0.03 \\
      $\tau$ & 67.5 & 32.5 & 41.2 & 58.8 \\
      \hline
      $w_0$ & 73.0 & 27.0 & \multirow{2}{*}{N/A} & \multirow{2}{*}{N/A} \\
      $w_a$ & 78.7 & 21.3 & &  \\
      \hline
      $\Omega_{m}$ & \multirow{2}{*}{N/A} & \multirow{2}{*}{N/A} & 0.9 & \cellcolor{redD}99.1 \\
      $S_8$ & & & 0.7 & \cellcolor{redD}99.3\\
      \hline
      $A_s e^{-2 \tau}$ & 6.1 & \cellcolor{redD}93.9 & 12.2 & 87.8 \\
      \hline 
    \end{tabular}
  \end{center}
  \label{table:parameter shifts}
\end{table}

Finally, we ask what implications are
on two of the most readily measured parameters by lensing surveys --- $\Omega_M$ and $S_8\equiv \sigma_8 (\Omega_M/0.3)^{0.5}$. Note that the values of these two parameters measured in lensing surveys and the CMB are typically interpreted within the context of the flat $\Lambda$CDM cosmological model. Therefore, to infer $\sigma_8$ from our set of simulated Horndeski data vectors, we now enforce a fit of modified gravity with a $\Lambda$CDM cosmology rather than $w_0 w_a$CDM. We thus fix $w_0 = -1$ and $w_a = 0$, and vary the six other parameters listed in Eq.~(\ref{eq:params}) to find the best-fit $\Lambda$CDM model. Then, we use \texttt{CAMB} to calculate the value of $\sigma_8$ and the corresponding $S_8$ for each best-fit $\Lambda$CDM model. 

\begin{figure}
    \centering
    \includegraphics[width=\linewidth]{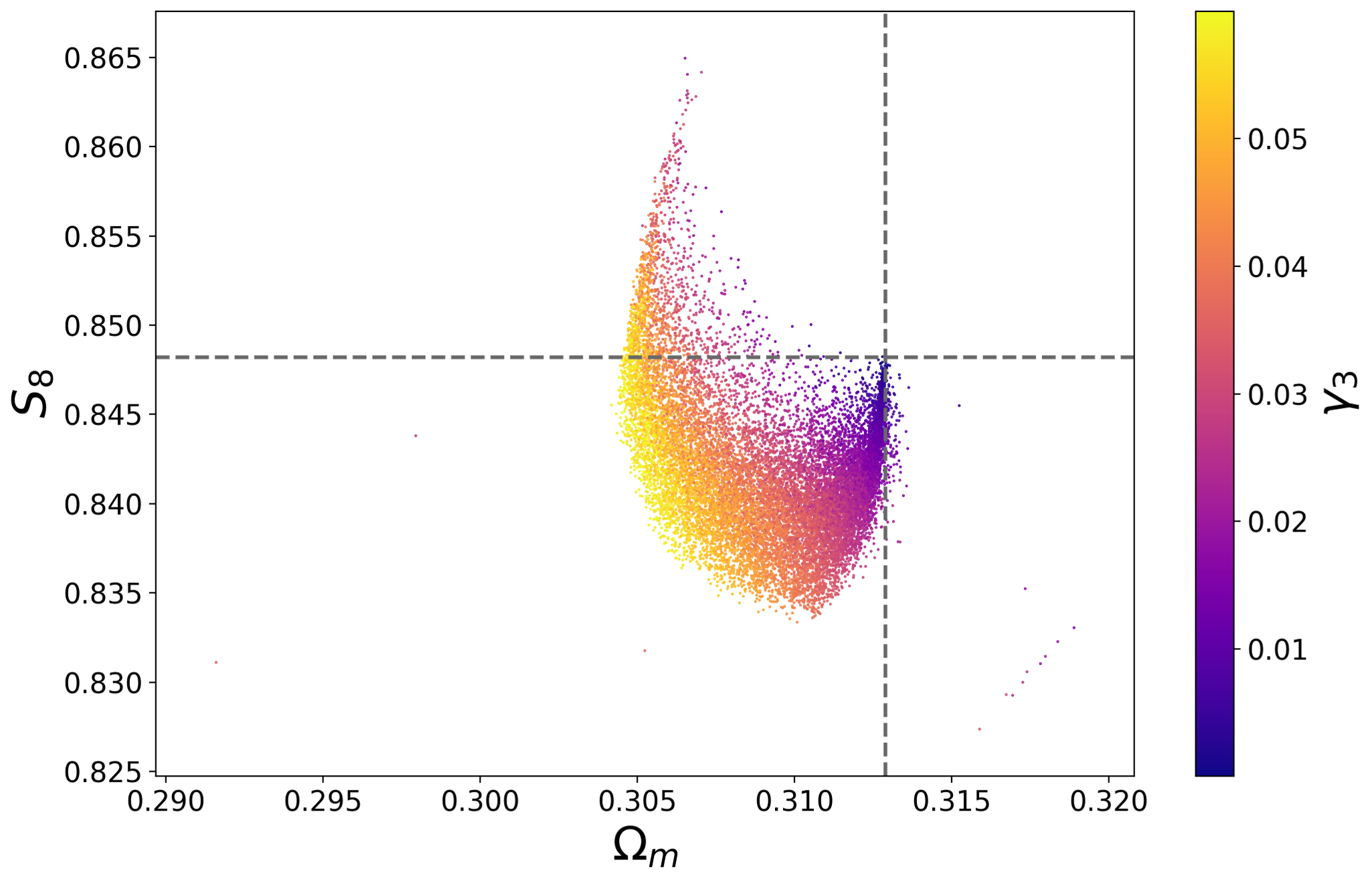}
    \caption{Best-fit values and projected 1D histograms of $\Omega_m$ and $S_8$ derived from fitting 16769 Horndeski data vectors with a $\Lambda$CDM cosmological model. Each point is colored by the $\gamma_3$ parameter for each Horndeski model as defined in Eq.~\ref{eq: gammas}.}
    \label{fig:omegaM S8}
\end{figure}

We plot $\Lambda$CDM's best-fit ($\Omega_M.S_8$) pair for each Horndeski model in Fig.~\ref{fig:omegaM S8}. Each point is colored by the $\gamma_3$ parameter (as defined in Eq.~\ref{eq: gammas}) for each Horndeski model we fitted to. As before, the cross-hairs denote the fiducial, input values of these parameters. In this case, we do not observe a particularly narrow region, or multiple branches, in the best-fit $\Omega_M-S_8$ plane. Rather, we see a near-universal shift to lower values of the best-fit $\Omega_M$, and also a preferential shift toward lower $S_8$. As the Horndeski model deviates more from general relativity as represented by larger values of $\gamma_3$, we observe a shift in $\Omega_m$ towards lower values. It is known that Horndeski models can generally accommodate both a larger and a smaller amplitude of structure formation relative to the standard model with the same background parameters. 
The results in Fig.~\ref{fig:omegaM S8} indicate that, for the range of Horndeski parameters that we investigated (see Eq.~\ref{eq:Horndeski-range}), these models near-universally show up as a smaller overall amplitude of mass fluctuations.

\begin{figure*}
    \centering
    \includegraphics[width=\textwidth]{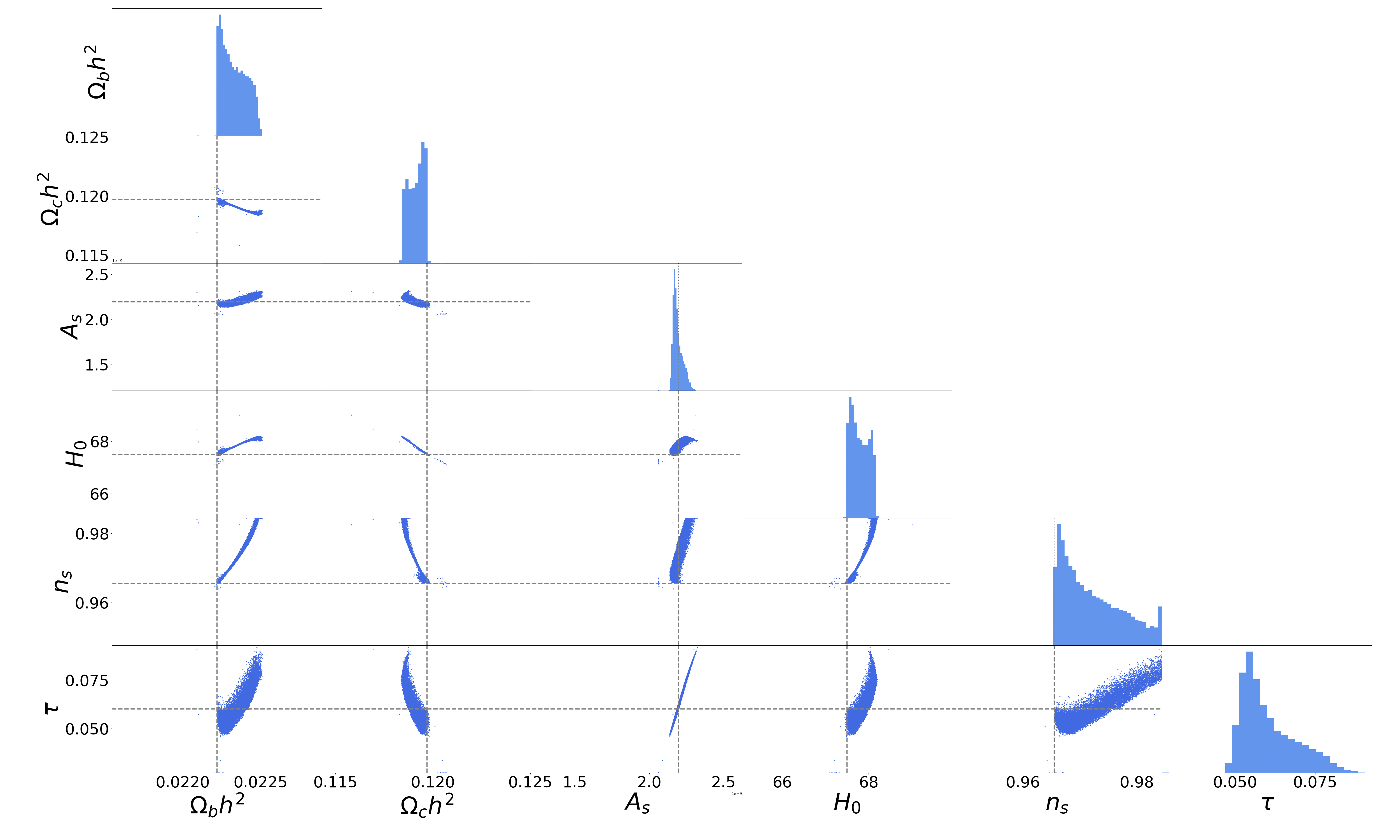}
    \caption{Best-fit values and histograms of cosmological parameters obtained from fitting to 16769 Horndeski models with a $\Lambda$CDM cosmology.}
    \label{fig:triangular LCDM}
\end{figure*}

We also investigated the biases that one would observe on all six base cosmological parameters when interpreting modified gravity with a $\Lambda$CDM cosmology. The results are displayed in Fig.~\ref{fig:triangular LCDM}, which contains all possible 2D planes and histograms of cosmological parameters. The grey crosshair again indicates the unbiased, fiducial value of a parameter. In every panel, each point represents a parameter's relative shift or bias resulting from misinterpreting one of the 16769 modified gravity models with dark energy. Here, we observe a shift towards a uniform direction among four of the six parameters, $\Omega_b h^2$, $\Omega_c h^2$, $H_0$ and $n_s$, which are listed in Table~\ref{table:parameter shifts}. The degenerate combination of $A_s e^{-2 \tau}$ also mostly shifts towards a value smaller than the fiducial one.

\section{Conclusion}\label{sec:concl}

In this work we address the question of how analyses that fit standard cosmological models (say \lcdm\ or $w_0w_a$CDM) to data may show hints of modified gravity. Assume for the moment that modified gravity is at work.  
In a realistic situation, it is entirely plausible that a standard, unmodified-gravity model is a good fit to the data, so that we cannot immediately rule it out and claim evidence for modified gravity. This scenario, however, will generally lead to shifts in the (standard-model, unmodified-gravity) parameter values  relative to their true values. And such shifts, interpreted together and in relation to other measurements in cosmology that depend on different kinds of data, may reveal the presence of modified gravity.

In this paper, we quantitatively investigate these parameter biases in scenarios when modified gravity is misinterpreted as a standard model. Specifically, we establish the link between modified-gravity models and shifts in the standard cosmological parameters. To scan through a broad range of modified-gravity model, we focus on the Horndeski universality class of models, whose phenomenological predictions (on linear scales) are produced by the code \texttt{EFTCAMB} \cite{Hu:2014oga}. Horndeski models allow a separate specification of the cosmological theory background and perturbations. For simplicity, we assume a cosmological-constant background for the Horndeski models (in agreement with the most recent cosmological data to date), and vary the perturbations, allowing the full freedom of Horndeski models. We fit these models with simulated future data consisting of CMB temperature and the polarization power spectra, BAO data, and type Ia supernova data. We restrict the analysis to only those Horndeski models whose simulated data vectors are well fit by the $w_0w_a$CDM model. In doing this we mimic a realistic situation where one would only proceed with the interpretation of model fits in scenarios where the goodness of fit passes some threshold. 

We report the best-fit values of the standard cosmological parameters for each Horndeski model that passes the aforementioned cuts. We find that the distribution of the best-fit values cover remarkably tight regions in the standard eight-dimensional parameter space (Fig.~\ref{fig:triangular}). These regions are largely linear, though on occasion carve out multi-pronged directions in the 2D parameter spaces. These tight correlations in standard parameter best-fits imply that even general classes of modified-gravity models register as specific deviations (from true values) in the unmodified-gravity parameters. This is good news; for example, a deviation in standard parameters that does \textit{not} lie in one of these directions would indicate that systematic errors, rather than modified gravity, may be the cause of such unexpected shifts. Hence it should be possible to spot such signatures of systematic errors in future data.

Focusing now on the equation-of-state parameter values that are best fits to Horndeski models, we find that, even though significant deviations in both $w_0$ and $w_a$ are allowed, they obey a tight mutual relation (Fig.~\ref{fig:w0wa}). Specifically, most Horndeski models are fit with an effective equation of state of $w(z_{\rm eff})\simeq -1$, evaluated at the effective redshift of $z_{\rm eff}=0.28$. This can be taken as a very generic prediction of the perturbations provided by the large class of modified-gravity models that we study, given a $\Lambda$CDM background as stipulated above. This prediction, along with those on all other parameters specified in Fig.~\ref{fig:triangular}, will be sharply tested using upcoming cosmological data.

We finally study the implications of our result to the currently much debated tension between constraints on the $S_8$ parameter obtained from lensing probes and CMB measurements. Assuming now the $\Lambda$CDM model (in which the $S_8$ tension is usually framed), we find that Horndeski models typically predict a lower $S_8$, and near-universally a lower $\Omega_M$, than the truth when the latter two are inferred assuming the $\Lambda$CDM model. Because the only direct probe of $S_8$ that we assumed was the CMB, this implies that CMB's $S_8$ value is preferentially low when Horndeski data are analyzed assuming the $\Lambda$CDM model. This should be compared to the prediction from applying the same pipeline to lensing data, something we plan to do in a future work.


\section{Acknowledgments}
We would like to thank Marco Raveri and Alessandra Silvestri for useful conversations, Anqi Chen for her initial collaboration, and Earl Lawrence for assistance on the usage of the emulator. We also thank Eric Linder for useful comments on an earlier draft of this paper. The work of EN and SW was supported in part by NSF grant 1813834 and NSF grant  PHY-1748958. They would also like to thank the Kavli Institute for Theoretical Physics and the Aspen Center for Physics for hospitality. YW and DH have been supported by NSF under contract AST-1812961. DH has additionally been supported by DOE under Contract No. DE-FG02-95ER40899, and he thanks the Humboldt Foundation for support via the Friedrich Wilhelm Bessel award.
We acknowledge the Syracuse University HTC Campus Grid and NSF award ACI-1341006 for the use of computing resources.

\section{Appendix: Fitting error}
\begin{figure*}
    \centering
    \includegraphics[width=\textwidth]{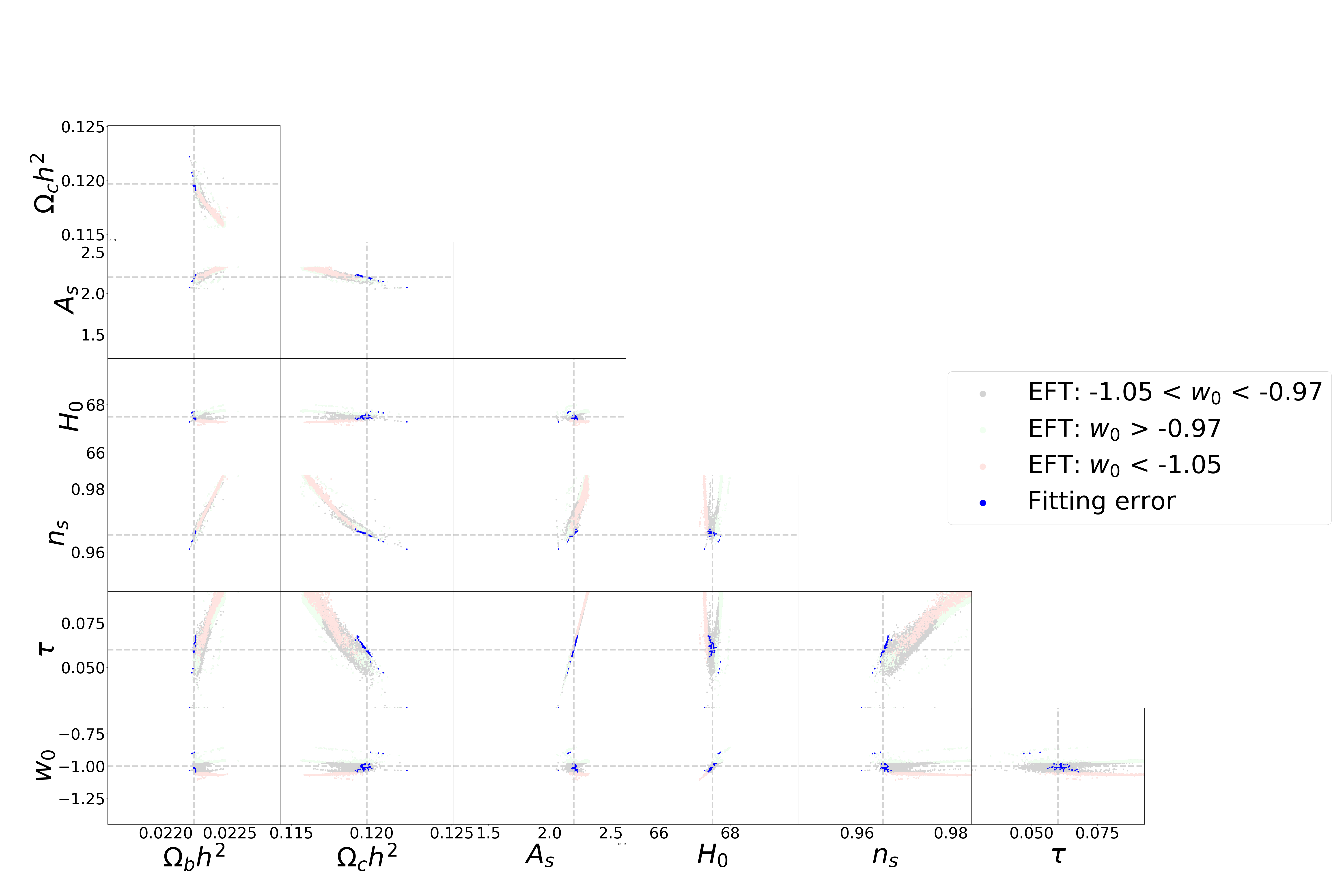}
    \caption{The extent of uncertainty in the process of finding best-fits as represented by the blue points, which are superimposed onto Fig.~\ref{fig:triangular}}.
    \label{fig:fitting error}
\end{figure*}

Here, we illustrate the extent of uncertainty in our process of finding best-fits. In each panel of Fig.~\ref{fig:fitting error}, there are 93 blue points, each generated from fitting the 8 standard cosmological parameters to the fiducial cosmology listed in Table~\ref{tab:parameters}. The dim light grey, green and red points in the background are the same as the corresponding points in Fig.~\ref{fig:triangular}, and in both figures they denote the best-fit parameter values to Horndeski data vectors. For a perfect fitting process, the blue points should all coincide with the grey crosshair, which indicates the fiducial values of each parameter. Our fitting error, as indicated by level of scatter among the blue points, is small compared to both the best-fits to Horndeski data vectors and the parameters' allowed ranges of variation. 

\newpage

\bibliography{refs}{}

\begin{thebibliography}{66}%
\makeatletter
\providecommand \@ifxundefined [1]{%
 \@ifx{#1\undefined}
}%
\providecommand \@ifnum [1]{%
 \ifnum #1\expandafter \@firstoftwo
 \else \expandafter \@secondoftwo
 \fi
}%
\providecommand \@ifx [1]{%
 \ifx #1\expandafter \@firstoftwo
 \else \expandafter \@secondoftwo
 \fi
}%
\providecommand \natexlab [1]{#1}%
\providecommand \enquote  [1]{``#1''}%
\providecommand \bibnamefont  [1]{#1}%
\providecommand \bibfnamefont [1]{#1}%
\providecommand \citenamefont [1]{#1}%
\providecommand \href@noop [0]{\@secondoftwo}%
\providecommand \href [0]{\begingroup \@sanitize@url \@href}%
\providecommand \@href[1]{\@@startlink{#1}\@@href}%
\providecommand \@@href[1]{\endgroup#1\@@endlink}%
\providecommand \@sanitize@url [0]{\catcode `\\12\catcode `\$12\catcode
  `\&12\catcode `\#12\catcode `\^12\catcode `\_12\catcode `\%12\relax}%
\providecommand \@@startlink[1]{}%
\providecommand \@@endlink[0]{}%
\providecommand \url  [0]{\begingroup\@sanitize@url \@url }%
\providecommand \@url [1]{\endgroup\@href {#1}{\urlprefix }}%
\providecommand \urlprefix  [0]{URL }%
\providecommand \Eprint [0]{\href }%
\providecommand \doibase [0]{http://dx.doi.org/}%
\providecommand \selectlanguage [0]{\@gobble}%
\providecommand \bibinfo  [0]{\@secondoftwo}%
\providecommand \bibfield  [0]{\@secondoftwo}%
\providecommand \translation [1]{[#1]}%
\providecommand \BibitemOpen [0]{}%
\providecommand \bibitemStop [0]{}%
\providecommand \bibitemNoStop [0]{.\EOS\space}%
\providecommand \EOS [0]{\spacefactor3000\relax}%
\providecommand \BibitemShut  [1]{\csname bibitem#1\endcsname}%
\let\auto@bib@innerbib\@empty
\bibitem [{\citenamefont {Frieman}\ \emph {et~al.}(2008)\citenamefont
  {Frieman}, \citenamefont {Turner},\ and\ \citenamefont
  {Huterer}}]{Frieman:2008sn}%
  \BibitemOpen
  \bibfield  {author} {\bibinfo {author} {\bibfnamefont {J.}~\bibnamefont
  {Frieman}}, \bibinfo {author} {\bibfnamefont {M.}~\bibnamefont {Turner}}, \
  and\ \bibinfo {author} {\bibfnamefont {D.}~\bibnamefont {Huterer}},\ }\href
  {\doibase 10.1146/annurev.astro.46.060407.145243} {\bibfield  {journal}
  {\bibinfo  {journal} {Ann. Rev. Astron. Astrophys.}\ }\textbf {\bibinfo
  {volume} {46}},\ \bibinfo {pages} {385} (\bibinfo {year} {2008})},\ \Eprint
  {http://arxiv.org/abs/0803.0982} {arXiv:0803.0982 [astro-ph]} \BibitemShut
  {NoStop}%
\bibitem [{\citenamefont {Huterer}\ and\ \citenamefont
  {Shafer}(2018)}]{Huterer:2017buf}%
  \BibitemOpen
  \bibfield  {author} {\bibinfo {author} {\bibfnamefont {D.}~\bibnamefont
  {Huterer}}\ and\ \bibinfo {author} {\bibfnamefont {D.~L.}\ \bibnamefont
  {Shafer}},\ }\href {\doibase 10.1088/1361-6633/aa997e} {\bibfield  {journal}
  {\bibinfo  {journal} {Rept. Prog. Phys.}\ }\textbf {\bibinfo {volume} {81}},\
  \bibinfo {pages} {016901} (\bibinfo {year} {2018})},\ \Eprint
  {http://arxiv.org/abs/1709.01091} {arXiv:1709.01091 [astro-ph.CO]}
  \BibitemShut {NoStop}%
\bibitem [{\citenamefont {Weinberg}(1989)}]{Weinberg:1988cp}%
  \BibitemOpen
  \bibfield  {author} {\bibinfo {author} {\bibfnamefont {S.}~\bibnamefont
  {Weinberg}},\ }\href {\doibase 10.1103/RevModPhys.61.1} {\bibfield  {journal}
  {\bibinfo  {journal} {Rev. Mod. Phys.}\ }\textbf {\bibinfo {volume} {61}},\
  \bibinfo {pages} {1} (\bibinfo {year} {1989})}\BibitemShut {NoStop}%
\bibitem [{\citenamefont {Carroll}(2001)}]{Carroll:2000fy}%
  \BibitemOpen
  \bibfield  {author} {\bibinfo {author} {\bibfnamefont {S.~M.}\ \bibnamefont
  {Carroll}},\ }\href {\doibase 10.12942/lrr-2001-1} {\bibfield  {journal}
  {\bibinfo  {journal} {Living Rev. Rel.}\ }\textbf {\bibinfo {volume} {4}},\
  \bibinfo {pages} {1} (\bibinfo {year} {2001})},\ \Eprint
  {http://arxiv.org/abs/astro-ph/0004075} {arXiv:astro-ph/0004075} \BibitemShut
  {NoStop}%
\bibitem [{\citenamefont {Copeland}\ \emph {et~al.}(2006)\citenamefont
  {Copeland}, \citenamefont {Sami},\ and\ \citenamefont
  {Tsujikawa}}]{Copeland2006}%
  \BibitemOpen
  \bibfield  {author} {\bibinfo {author} {\bibfnamefont {E.~J.}\ \bibnamefont
  {Copeland}}, \bibinfo {author} {\bibfnamefont {M.}~\bibnamefont {Sami}}, \
  and\ \bibinfo {author} {\bibfnamefont {S.}~\bibnamefont {Tsujikawa}},\ }\href
  {\doibase 10.1142/S021827180600942X} {\bibfield  {journal} {\bibinfo
  {journal} {International Journal of Modern Physics D}\ }\textbf {\bibinfo
  {volume} {15}},\ \bibinfo {pages} {1753} (\bibinfo {year} {2006})},\ \Eprint
  {http://arxiv.org/abs/0603057} {arXiv:0603057 [hep-th]} \BibitemShut
  {NoStop}%
\bibitem [{\citenamefont {Clifton}\ \emph {et~al.}(2012)\citenamefont
  {Clifton}, \citenamefont {Ferreira}, \citenamefont {Padilla},\ and\
  \citenamefont {Skordis}}]{Clifton2012}%
  \BibitemOpen
  \bibfield  {author} {\bibinfo {author} {\bibfnamefont {T.}~\bibnamefont
  {Clifton}}, \bibinfo {author} {\bibfnamefont {P.~G.}\ \bibnamefont
  {Ferreira}}, \bibinfo {author} {\bibfnamefont {A.}~\bibnamefont {Padilla}}, \
  and\ \bibinfo {author} {\bibfnamefont {C.}~\bibnamefont {Skordis}},\ }\href
  {\doibase 10.1016/j.physrep.2012.01.001} {\bibfield  {journal} {\bibinfo
  {journal} {Physics Reports}\ }\textbf {\bibinfo {volume} {513}},\ \bibinfo
  {pages} {1} (\bibinfo {year} {2012})},\ \Eprint
  {http://arxiv.org/abs/1106.2476} {arXiv:1106.2476} \BibitemShut {NoStop}%
\bibitem [{\citenamefont {Joyce}\ \emph {et~al.}(2015)\citenamefont {Joyce},
  \citenamefont {Jain}, \citenamefont {Khoury},\ and\ \citenamefont
  {Trodden}}]{Joyce2015}%
  \BibitemOpen
  \bibfield  {author} {\bibinfo {author} {\bibfnamefont {A.}~\bibnamefont
  {Joyce}}, \bibinfo {author} {\bibfnamefont {B.}~\bibnamefont {Jain}},
  \bibinfo {author} {\bibfnamefont {J.}~\bibnamefont {Khoury}}, \ and\ \bibinfo
  {author} {\bibfnamefont {M.}~\bibnamefont {Trodden}},\ }\href {\doibase
  10.1016/j.physrep.2014.12.002} {\bibfield  {journal} {\bibinfo  {journal}
  {Physics Reports}\ }\textbf {\bibinfo {volume} {568}},\ \bibinfo {pages} {1}
  (\bibinfo {year} {2015})},\ \Eprint {http://arxiv.org/abs/1407.0059}
  {arXiv:1407.0059} \BibitemShut {NoStop}%
\bibitem [{\citenamefont {Silvestri}\ and\ \citenamefont
  {Trodden}(2009)}]{Silvestri2009}%
  \BibitemOpen
  \bibfield  {author} {\bibinfo {author} {\bibfnamefont {A.}~\bibnamefont
  {Silvestri}}\ and\ \bibinfo {author} {\bibfnamefont {M.}~\bibnamefont
  {Trodden}},\ }\href {\doibase 10.1088/0034-4885/72/9/096901} {\bibfield
  {journal} {\bibinfo  {journal} {Reports on Progress in Physics}\ }\textbf
  {\bibinfo {volume} {72}} (\bibinfo {year} {2009}),\
  10.1088/0034-4885/72/9/096901},\ \Eprint {http://arxiv.org/abs/0904.0024}
  {arXiv:0904.0024} \BibitemShut {NoStop}%
\bibitem [{\citenamefont {Arnold}\ \emph {et~al.}(2019)\citenamefont {Arnold},
  \citenamefont {Leo},\ and\ \citenamefont {Li}}]{Arnold2019}%
  \BibitemOpen
  \bibfield  {author} {\bibinfo {author} {\bibfnamefont {C.}~\bibnamefont
  {Arnold}}, \bibinfo {author} {\bibfnamefont {M.}~\bibnamefont {Leo}}, \ and\
  \bibinfo {author} {\bibfnamefont {B.}~\bibnamefont {Li}},\ }\href {\doibase
  10.1038/s41550-019-0823-y} {\bibfield  {journal} {\bibinfo  {journal} {Nature
  Astronomy}\ }\textbf {\bibinfo {volume} {3}},\ \bibinfo {pages} {945}
  (\bibinfo {year} {2019})},\ \Eprint {http://arxiv.org/abs/1907.02977}
  {arXiv:1907.02977} \BibitemShut {NoStop}%
\bibitem [{\citenamefont {Mitchell}\ \emph {et~al.}(2021)\citenamefont
  {Mitchell}, \citenamefont {Arnold},\ and\ \citenamefont
  {Li}}]{Mitchell:2021uzh}%
  \BibitemOpen
  \bibfield  {author} {\bibinfo {author} {\bibfnamefont {M.~A.}\ \bibnamefont
  {Mitchell}}, \bibinfo {author} {\bibfnamefont {C.}~\bibnamefont {Arnold}}, \
  and\ \bibinfo {author} {\bibfnamefont {B.}~\bibnamefont {Li}},\ }\href@noop
  {} {\bibfield  {journal} {\bibinfo  {journal} {Mon. Not. Roy. Astron. Soc.}\
  }\textbf {\bibinfo {volume} {508}},\ \bibinfo {pages} {4157} (\bibinfo {year}
  {2021})},\ \Eprint {http://arxiv.org/abs/2107.14224} {arXiv:2107.14224
  [astro-ph.CO]} \BibitemShut {NoStop}%
\bibitem [{\citenamefont {Turner}\ and\ \citenamefont
  {White}(1997)}]{Turner:1997npq}%
  \BibitemOpen
  \bibfield  {author} {\bibinfo {author} {\bibfnamefont {M.~S.}\ \bibnamefont
  {Turner}}\ and\ \bibinfo {author} {\bibfnamefont {M.~J.}\ \bibnamefont
  {White}},\ }\href {\doibase 10.1103/PhysRevD.56.R4439} {\bibfield  {journal}
  {\bibinfo  {journal} {Phys. Rev. D}\ }\textbf {\bibinfo {volume} {56}},\
  \bibinfo {pages} {R4439} (\bibinfo {year} {1997})},\ \Eprint
  {http://arxiv.org/abs/astro-ph/9701138} {arXiv:astro-ph/9701138} \BibitemShut
  {NoStop}%
\bibitem [{\citenamefont {Linder}(2003)}]{Linder2003}%
  \BibitemOpen
  \bibfield  {author} {\bibinfo {author} {\bibfnamefont {E.~V.}\ \bibnamefont
  {Linder}},\ }\href {\doibase 10.1103/PhysRevLett.90.091301} {\bibfield
  {journal} {\bibinfo  {journal} {Physical Review Letters}\ }\textbf {\bibinfo
  {volume} {90}},\ \bibinfo {pages} {4} (\bibinfo {year} {2003})},\ \Eprint
  {http://arxiv.org/abs/0208512} {arXiv:0208512 [astro-ph]} \BibitemShut
  {NoStop}%
\bibitem [{\citenamefont {Zhang}\ \emph {et~al.}(2007)\citenamefont {Zhang},
  \citenamefont {Liguori}, \citenamefont {Bean},\ and\ \citenamefont
  {Dodelson}}]{Zhang:2007nk}%
  \BibitemOpen
  \bibfield  {author} {\bibinfo {author} {\bibfnamefont {P.}~\bibnamefont
  {Zhang}}, \bibinfo {author} {\bibfnamefont {M.}~\bibnamefont {Liguori}},
  \bibinfo {author} {\bibfnamefont {R.}~\bibnamefont {Bean}}, \ and\ \bibinfo
  {author} {\bibfnamefont {S.}~\bibnamefont {Dodelson}},\ }\href {\doibase
  10.1103/PhysRevLett.99.141302} {\bibfield  {journal} {\bibinfo  {journal}
  {Phys. Rev. Lett.}\ }\textbf {\bibinfo {volume} {99}},\ \bibinfo {pages}
  {141302} (\bibinfo {year} {2007})},\ \Eprint {http://arxiv.org/abs/0704.1932}
  {arXiv:0704.1932 [astro-ph]} \BibitemShut {NoStop}%
\bibitem [{\citenamefont {Daniel}\ and\ \citenamefont
  {Linder}(2013)}]{Daniel:2012kn}%
  \BibitemOpen
  \bibfield  {author} {\bibinfo {author} {\bibfnamefont {S.~F.}\ \bibnamefont
  {Daniel}}\ and\ \bibinfo {author} {\bibfnamefont {E.~V.}\ \bibnamefont
  {Linder}},\ }\href {\doibase 10.1088/1475-7516/2013/02/007} {\bibfield
  {journal} {\bibinfo  {journal} {JCAP}\ }\textbf {\bibinfo {volume} {1302}},\
  \bibinfo {pages} {007} (\bibinfo {year} {2013})},\ \Eprint
  {http://arxiv.org/abs/1212.0009} {arXiv:1212.0009 [astro-ph.CO]} \BibitemShut
  {NoStop}%
\bibitem [{\citenamefont {Pogosian}\ and\ \citenamefont
  {Silvestri}(2016)}]{Pogosian:2016pwr}%
  \BibitemOpen
  \bibfield  {author} {\bibinfo {author} {\bibfnamefont {L.}~\bibnamefont
  {Pogosian}}\ and\ \bibinfo {author} {\bibfnamefont {A.}~\bibnamefont
  {Silvestri}},\ }\href {\doibase 10.1103/PhysRevD.94.104014} {\bibfield
  {journal} {\bibinfo  {journal} {Phys. Rev. D}\ }\textbf {\bibinfo {volume}
  {94}},\ \bibinfo {pages} {104014} (\bibinfo {year} {2016})},\ \Eprint
  {http://arxiv.org/abs/1606.05339} {arXiv:1606.05339 [astro-ph.CO]}
  \BibitemShut {NoStop}%
\bibitem [{\citenamefont {Zhang}(2006)}]{Zhang:2005vt}%
  \BibitemOpen
  \bibfield  {author} {\bibinfo {author} {\bibfnamefont {P.}~\bibnamefont
  {Zhang}},\ }\href {\doibase 10.1103/PhysRevD.73.123504} {\bibfield  {journal}
  {\bibinfo  {journal} {Phys. Rev.}\ }\textbf {\bibinfo {volume} {D73}},\
  \bibinfo {pages} {123504} (\bibinfo {year} {2006})},\ \Eprint
  {http://arxiv.org/abs/astro-ph/0511218} {arXiv:astro-ph/0511218 [astro-ph]}
  \BibitemShut {NoStop}%
\bibitem [{\citenamefont {Caldwell}\ \emph {et~al.}(2007)\citenamefont
  {Caldwell}, \citenamefont {Cooray},\ and\ \citenamefont
  {Melchiorri}}]{Caldwell:2007cw}%
  \BibitemOpen
  \bibfield  {author} {\bibinfo {author} {\bibfnamefont {R.}~\bibnamefont
  {Caldwell}}, \bibinfo {author} {\bibfnamefont {A.}~\bibnamefont {Cooray}}, \
  and\ \bibinfo {author} {\bibfnamefont {A.}~\bibnamefont {Melchiorri}},\
  }\href {\doibase 10.1103/PhysRevD.76.023507} {\bibfield  {journal} {\bibinfo
  {journal} {Phys. Rev.}\ }\textbf {\bibinfo {volume} {D76}},\ \bibinfo {pages}
  {023507} (\bibinfo {year} {2007})},\ \Eprint
  {http://arxiv.org/abs/astro-ph/0703375} {arXiv:astro-ph/0703375 [ASTRO-PH]}
  \BibitemShut {NoStop}%
\bibitem [{\citenamefont {Guzik}\ \emph {et~al.}(2010)\citenamefont {Guzik},
  \citenamefont {Jain},\ and\ \citenamefont {Takada}}]{Guzik:2009cm}%
  \BibitemOpen
  \bibfield  {author} {\bibinfo {author} {\bibfnamefont {J.}~\bibnamefont
  {Guzik}}, \bibinfo {author} {\bibfnamefont {B.}~\bibnamefont {Jain}}, \ and\
  \bibinfo {author} {\bibfnamefont {M.}~\bibnamefont {Takada}},\ }\href
  {\doibase 10.1103/PhysRevD.81.023503} {\bibfield  {journal} {\bibinfo
  {journal} {Phys. Rev.}\ }\textbf {\bibinfo {volume} {D81}},\ \bibinfo {pages}
  {023503} (\bibinfo {year} {2010})},\ \Eprint {http://arxiv.org/abs/0906.2221}
  {arXiv:0906.2221 [astro-ph.CO]} \BibitemShut {NoStop}%
\bibitem [{\citenamefont {Bean}\ and\ \citenamefont
  {Tangmatitham}(2010)}]{Bean:2010zq}%
  \BibitemOpen
  \bibfield  {author} {\bibinfo {author} {\bibfnamefont {R.}~\bibnamefont
  {Bean}}\ and\ \bibinfo {author} {\bibfnamefont {M.}~\bibnamefont
  {Tangmatitham}},\ }\href {\doibase 10.1103/PhysRevD.81.083534} {\bibfield
  {journal} {\bibinfo  {journal} {Phys. Rev.}\ }\textbf {\bibinfo {volume}
  {D81}},\ \bibinfo {pages} {083534} (\bibinfo {year} {2010})},\ \Eprint
  {http://arxiv.org/abs/1002.4197} {arXiv:1002.4197 [astro-ph.CO]} \BibitemShut
  {NoStop}%
\bibitem [{\citenamefont {Zhao}\ \emph {et~al.}(2010)\citenamefont {Zhao},
  \citenamefont {Giannantonio}, \citenamefont {Pogosian}, \citenamefont
  {Silvestri}, \citenamefont {Bacon}, \citenamefont {Koyama}, \citenamefont
  {Nichol},\ and\ \citenamefont {Song}}]{Zhao:2010dz}%
  \BibitemOpen
  \bibfield  {author} {\bibinfo {author} {\bibfnamefont {G.-B.}\ \bibnamefont
  {Zhao}}, \bibinfo {author} {\bibfnamefont {T.}~\bibnamefont {Giannantonio}},
  \bibinfo {author} {\bibfnamefont {L.}~\bibnamefont {Pogosian}}, \bibinfo
  {author} {\bibfnamefont {A.}~\bibnamefont {Silvestri}}, \bibinfo {author}
  {\bibfnamefont {D.~J.}\ \bibnamefont {Bacon}}, \bibinfo {author}
  {\bibfnamefont {K.}~\bibnamefont {Koyama}}, \bibinfo {author} {\bibfnamefont
  {R.~C.}\ \bibnamefont {Nichol}}, \ and\ \bibinfo {author} {\bibfnamefont
  {Y.-S.}\ \bibnamefont {Song}},\ }\href {\doibase 10.1103/PhysRevD.81.103510}
  {\bibfield  {journal} {\bibinfo  {journal} {Phys. Rev.}\ }\textbf {\bibinfo
  {volume} {D81}},\ \bibinfo {pages} {103510} (\bibinfo {year} {2010})},\
  \Eprint {http://arxiv.org/abs/1003.0001} {arXiv:1003.0001 [astro-ph.CO]}
  \BibitemShut {NoStop}%
\bibitem [{\citenamefont {Reyes}\ \emph {et~al.}(2010)\citenamefont {Reyes},
  \citenamefont {Mandelbaum}, \citenamefont {Seljak}, \citenamefont {Baldauf},
  \citenamefont {Gunn}, \citenamefont {Lombriser},\ and\ \citenamefont
  {Smith}}]{Reyes:2010tr}%
  \BibitemOpen
  \bibfield  {author} {\bibinfo {author} {\bibfnamefont {R.}~\bibnamefont
  {Reyes}}, \bibinfo {author} {\bibfnamefont {R.}~\bibnamefont {Mandelbaum}},
  \bibinfo {author} {\bibfnamefont {U.}~\bibnamefont {Seljak}}, \bibinfo
  {author} {\bibfnamefont {T.}~\bibnamefont {Baldauf}}, \bibinfo {author}
  {\bibfnamefont {J.~E.}\ \bibnamefont {Gunn}}, \bibinfo {author}
  {\bibfnamefont {L.}~\bibnamefont {Lombriser}}, \ and\ \bibinfo {author}
  {\bibfnamefont {R.~E.}\ \bibnamefont {Smith}},\ }\href {\doibase
  10.1038/nature08857} {\bibfield  {journal} {\bibinfo  {journal} {Nature}\
  }\textbf {\bibinfo {volume} {464}},\ \bibinfo {pages} {256} (\bibinfo {year}
  {2010})},\ \Eprint {http://arxiv.org/abs/1003.2185} {arXiv:1003.2185
  [astro-ph.CO]} \BibitemShut {NoStop}%
\bibitem [{\citenamefont {Daniel}\ \emph {et~al.}(2010)\citenamefont {Daniel},
  \citenamefont {Linder}, \citenamefont {Smith}, \citenamefont {Caldwell},
  \citenamefont {Cooray}, \citenamefont {Leauthaud},\ and\ \citenamefont
  {Lombriser}}]{Daniel:2010ky}%
  \BibitemOpen
  \bibfield  {author} {\bibinfo {author} {\bibfnamefont {S.~F.}\ \bibnamefont
  {Daniel}}, \bibinfo {author} {\bibfnamefont {E.~V.}\ \bibnamefont {Linder}},
  \bibinfo {author} {\bibfnamefont {T.~L.}\ \bibnamefont {Smith}}, \bibinfo
  {author} {\bibfnamefont {R.~R.}\ \bibnamefont {Caldwell}}, \bibinfo {author}
  {\bibfnamefont {A.}~\bibnamefont {Cooray}}, \bibinfo {author} {\bibfnamefont
  {A.}~\bibnamefont {Leauthaud}}, \ and\ \bibinfo {author} {\bibfnamefont
  {L.}~\bibnamefont {Lombriser}},\ }\href {\doibase 10.1103/PhysRevD.81.123508}
  {\bibfield  {journal} {\bibinfo  {journal} {Phys. Rev.}\ }\textbf {\bibinfo
  {volume} {D81}},\ \bibinfo {pages} {123508} (\bibinfo {year} {2010})},\
  \Eprint {http://arxiv.org/abs/1002.1962} {arXiv:1002.1962 [astro-ph.CO]}
  \BibitemShut {NoStop}%
\bibitem [{\citenamefont {Daniel}\ and\ \citenamefont
  {Linder}(2010)}]{Daniel:2010yt}%
  \BibitemOpen
  \bibfield  {author} {\bibinfo {author} {\bibfnamefont {S.~F.}\ \bibnamefont
  {Daniel}}\ and\ \bibinfo {author} {\bibfnamefont {E.~V.}\ \bibnamefont
  {Linder}},\ }\href {\doibase 10.1103/PhysRevD.82.103523} {\bibfield
  {journal} {\bibinfo  {journal} {Phys. Rev. D}\ }\textbf {\bibinfo {volume}
  {82}},\ \bibinfo {pages} {103523} (\bibinfo {year} {2010})},\ \Eprint
  {http://arxiv.org/abs/1008.0397} {arXiv:1008.0397 [astro-ph.CO]} \BibitemShut
  {NoStop}%
\bibitem [{\citenamefont {Zhao}\ \emph {et~al.}(2012)\citenamefont {Zhao},
  \citenamefont {Crittenden}, \citenamefont {Pogosian},\ and\ \citenamefont
  {Zhang}}]{Zhao:2012aw}%
  \BibitemOpen
  \bibfield  {author} {\bibinfo {author} {\bibfnamefont {G.-B.}\ \bibnamefont
  {Zhao}}, \bibinfo {author} {\bibfnamefont {R.~G.}\ \bibnamefont
  {Crittenden}}, \bibinfo {author} {\bibfnamefont {L.}~\bibnamefont
  {Pogosian}}, \ and\ \bibinfo {author} {\bibfnamefont {X.}~\bibnamefont
  {Zhang}},\ }\href {\doibase 10.1103/PhysRevLett.109.171301} {\bibfield
  {journal} {\bibinfo  {journal} {Phys. Rev. Lett.}\ }\textbf {\bibinfo
  {volume} {109}},\ \bibinfo {pages} {171301} (\bibinfo {year} {2012})},\
  \Eprint {http://arxiv.org/abs/1207.3804} {arXiv:1207.3804 [astro-ph.CO]}
  \BibitemShut {NoStop}%
\bibitem [{\citenamefont {Raveri}\ \emph {et~al.}(2014)\citenamefont {Raveri},
  \citenamefont {Hu}, \citenamefont {Frusciante},\ and\ \citenamefont
  {Silvestri}}]{Raveri:2014cka}%
  \BibitemOpen
  \bibfield  {author} {\bibinfo {author} {\bibfnamefont {M.}~\bibnamefont
  {Raveri}}, \bibinfo {author} {\bibfnamefont {B.}~\bibnamefont {Hu}}, \bibinfo
  {author} {\bibfnamefont {N.}~\bibnamefont {Frusciante}}, \ and\ \bibinfo
  {author} {\bibfnamefont {A.}~\bibnamefont {Silvestri}},\ }\href {\doibase
  10.1103/PhysRevD.90.043513} {\bibfield  {journal} {\bibinfo  {journal} {Phys.
  Rev. D}\ }\textbf {\bibinfo {volume} {90}},\ \bibinfo {pages} {043513}
  (\bibinfo {year} {2014})},\ \Eprint {http://arxiv.org/abs/1405.1022}
  {arXiv:1405.1022 [astro-ph.CO]} \BibitemShut {NoStop}%
\bibitem [{\citenamefont {Bellini}\ \emph {et~al.}(2016)\citenamefont
  {Bellini}, \citenamefont {Cuesta}, \citenamefont {Jimenez},\ and\
  \citenamefont {Verde}}]{Bellini:2015xja}%
  \BibitemOpen
  \bibfield  {author} {\bibinfo {author} {\bibfnamefont {E.}~\bibnamefont
  {Bellini}}, \bibinfo {author} {\bibfnamefont {A.~J.}\ \bibnamefont {Cuesta}},
  \bibinfo {author} {\bibfnamefont {R.}~\bibnamefont {Jimenez}}, \ and\
  \bibinfo {author} {\bibfnamefont {L.}~\bibnamefont {Verde}},\ }\href
  {\doibase 10.1088/1475-7516/2016/06/E01} {\bibfield  {journal} {\bibinfo
  {journal} {JCAP}\ }\textbf {\bibinfo {volume} {02}},\ \bibinfo {pages} {053}
  (\bibinfo {year} {2016})},\ \bibinfo {note} {[Erratum: JCAP 06, E01
  (2016)]},\ \Eprint {http://arxiv.org/abs/1509.07816} {arXiv:1509.07816
  [astro-ph.CO]} \BibitemShut {NoStop}%
\bibitem [{\citenamefont {Hojjati}\ \emph {et~al.}(2016)\citenamefont
  {Hojjati}, \citenamefont {Plahn}, \citenamefont {Zucca}, \citenamefont
  {Pogosian}, \citenamefont {Brax}, \citenamefont {Davis},\ and\ \citenamefont
  {Zhao}}]{Hojjati:2015ojt}%
  \BibitemOpen
  \bibfield  {author} {\bibinfo {author} {\bibfnamefont {A.}~\bibnamefont
  {Hojjati}}, \bibinfo {author} {\bibfnamefont {A.}~\bibnamefont {Plahn}},
  \bibinfo {author} {\bibfnamefont {A.}~\bibnamefont {Zucca}}, \bibinfo
  {author} {\bibfnamefont {L.}~\bibnamefont {Pogosian}}, \bibinfo {author}
  {\bibfnamefont {P.}~\bibnamefont {Brax}}, \bibinfo {author} {\bibfnamefont
  {A.-C.}\ \bibnamefont {Davis}}, \ and\ \bibinfo {author} {\bibfnamefont
  {G.-B.}\ \bibnamefont {Zhao}},\ }\href {\doibase 10.1103/PhysRevD.93.043531}
  {\bibfield  {journal} {\bibinfo  {journal} {Phys. Rev.}\ }\textbf {\bibinfo
  {volume} {D93}},\ \bibinfo {pages} {043531} (\bibinfo {year} {2016})},\
  \Eprint {http://arxiv.org/abs/1511.05962} {arXiv:1511.05962 [astro-ph.CO]}
  \BibitemShut {NoStop}%
\bibitem [{\citenamefont {Salvatelli}\ \emph {et~al.}(2016)\citenamefont
  {Salvatelli}, \citenamefont {Piazza},\ and\ \citenamefont
  {Marinoni}}]{Salvatelli:2016mgy}%
  \BibitemOpen
  \bibfield  {author} {\bibinfo {author} {\bibfnamefont {V.}~\bibnamefont
  {Salvatelli}}, \bibinfo {author} {\bibfnamefont {F.}~\bibnamefont {Piazza}},
  \ and\ \bibinfo {author} {\bibfnamefont {C.}~\bibnamefont {Marinoni}},\
  }\href {\doibase 10.1088/1475-7516/2016/09/027} {\bibfield  {journal}
  {\bibinfo  {journal} {JCAP}\ }\textbf {\bibinfo {volume} {1609}},\ \bibinfo
  {pages} {027} (\bibinfo {year} {2016})},\ \Eprint
  {http://arxiv.org/abs/1602.08283} {arXiv:1602.08283 [astro-ph.CO]}
  \BibitemShut {NoStop}%
\bibitem [{\citenamefont {Joudaki}\ \emph {et~al.}(2017)\citenamefont {Joudaki}
  \emph {et~al.}}]{Joudaki:2016kym}%
  \BibitemOpen
  \bibfield  {author} {\bibinfo {author} {\bibfnamefont {S.}~\bibnamefont
  {Joudaki}} \emph {et~al.},\ }\href {\doibase 10.1093/mnras/stx998} {\bibfield
   {journal} {\bibinfo  {journal} {Mon. Not. Roy. Astron. Soc.}\ }\textbf
  {\bibinfo {volume} {471}},\ \bibinfo {pages} {1259} (\bibinfo {year}
  {2017})},\ \Eprint {http://arxiv.org/abs/1610.04606} {arXiv:1610.04606
  [astro-ph.CO]} \BibitemShut {NoStop}%
\bibitem [{\citenamefont {Mueller}\ \emph {et~al.}(2018)\citenamefont
  {Mueller}, \citenamefont {Percival}, \citenamefont {Linder}, \citenamefont
  {Alam}, \citenamefont {Zhao}, \citenamefont {Sánchez}, \citenamefont
  {Beutler},\ and\ \citenamefont {Brinkmann}}]{Mueller:2016kpu}%
  \BibitemOpen
  \bibfield  {author} {\bibinfo {author} {\bibfnamefont {E.-M.}\ \bibnamefont
  {Mueller}}, \bibinfo {author} {\bibfnamefont {W.}~\bibnamefont {Percival}},
  \bibinfo {author} {\bibfnamefont {E.}~\bibnamefont {Linder}}, \bibinfo
  {author} {\bibfnamefont {S.}~\bibnamefont {Alam}}, \bibinfo {author}
  {\bibfnamefont {G.-B.}\ \bibnamefont {Zhao}}, \bibinfo {author}
  {\bibfnamefont {A.~G.}\ \bibnamefont {Sánchez}}, \bibinfo {author}
  {\bibfnamefont {F.}~\bibnamefont {Beutler}}, \ and\ \bibinfo {author}
  {\bibfnamefont {J.}~\bibnamefont {Brinkmann}},\ }\href {\doibase
  10.1093/mnras/stx3232} {\bibfield  {journal} {\bibinfo  {journal} {Mon. Not.
  Roy. Astron. Soc.}\ }\textbf {\bibinfo {volume} {475}},\ \bibinfo {pages}
  {2122} (\bibinfo {year} {2018})},\ \Eprint {http://arxiv.org/abs/1612.00812}
  {arXiv:1612.00812 [astro-ph.CO]} \BibitemShut {NoStop}%
\bibitem [{\citenamefont {Zhao}\ \emph {et~al.}(2017)\citenamefont {Zhao} \emph
  {et~al.}}]{Zhao:2017cud}%
  \BibitemOpen
  \bibfield  {author} {\bibinfo {author} {\bibfnamefont {G.-B.}\ \bibnamefont
  {Zhao}} \emph {et~al.},\ }\href {\doibase 10.1038/s41550-017-0216-z}
  {\bibfield  {journal} {\bibinfo  {journal} {Nat. Astron.}\ }\textbf {\bibinfo
  {volume} {1}},\ \bibinfo {pages} {627} (\bibinfo {year} {2017})},\ \Eprint
  {http://arxiv.org/abs/1701.08165} {arXiv:1701.08165 [astro-ph.CO]}
  \BibitemShut {NoStop}%
\bibitem [{\citenamefont {Amon}\ \emph {et~al.}(2018)\citenamefont {Amon} \emph
  {et~al.}}]{Amon:2017lia}%
  \BibitemOpen
  \bibfield  {author} {\bibinfo {author} {\bibfnamefont {A.}~\bibnamefont
  {Amon}} \emph {et~al.},\ }\href@noop {} {\bibfield  {journal} {\bibinfo
  {journal} {Mon. Not. Roy. Astron. Soc.}\ }\textbf {\bibinfo {volume} {479}},\
  \bibinfo {pages} {3422} (\bibinfo {year} {2018})},\ \Eprint
  {http://arxiv.org/abs/1711.10999} {arXiv:1711.10999 [astro-ph.CO]}
  \BibitemShut {NoStop}%
\bibitem [{\citenamefont {Aghanim}\ \emph
  {et~al.}(2020{\natexlab{a}})\citenamefont {Aghanim} \emph
  {et~al.}}]{Aghanim:2018eyx}%
  \BibitemOpen
  \bibfield  {author} {\bibinfo {author} {\bibfnamefont {N.}~\bibnamefont
  {Aghanim}} \emph {et~al.} (\bibinfo {collaboration} {Planck}),\ }\href
  {\doibase 10.1051/0004-6361/201833910} {\bibfield  {journal} {\bibinfo
  {journal} {Astron. Astrophys.}\ }\textbf {\bibinfo {volume} {641}},\ \bibinfo
  {pages} {A6} (\bibinfo {year} {2020}{\natexlab{a}})},\ \Eprint
  {http://arxiv.org/abs/1807.06209} {arXiv:1807.06209 [astro-ph.CO]}
  \BibitemShut {NoStop}%
\bibitem [{\citenamefont {Abbott}\ \emph {et~al.}(2019)\citenamefont {Abbott}
  \emph {et~al.}}]{DES:2018ufa}%
  \BibitemOpen
  \bibfield  {author} {\bibinfo {author} {\bibfnamefont {T.~M.~C.}\
  \bibnamefont {Abbott}} \emph {et~al.} (\bibinfo {collaboration} {DES}),\
  }\href {\doibase 10.1103/PhysRevD.99.123505} {\bibfield  {journal} {\bibinfo
  {journal} {Phys. Rev. D}\ }\textbf {\bibinfo {volume} {99}},\ \bibinfo
  {pages} {123505} (\bibinfo {year} {2019})},\ \Eprint
  {http://arxiv.org/abs/1810.02499} {arXiv:1810.02499 [astro-ph.CO]}
  \BibitemShut {NoStop}%
\bibitem [{\citenamefont {Noller}\ and\ \citenamefont
  {Nicola}(2019)}]{Noller:2018wyv}%
  \BibitemOpen
  \bibfield  {author} {\bibinfo {author} {\bibfnamefont {J.}~\bibnamefont
  {Noller}}\ and\ \bibinfo {author} {\bibfnamefont {A.}~\bibnamefont
  {Nicola}},\ }\href {\doibase 10.1103/PhysRevD.99.103502} {\bibfield
  {journal} {\bibinfo  {journal} {Phys. Rev. D}\ }\textbf {\bibinfo {volume}
  {99}},\ \bibinfo {pages} {103502} (\bibinfo {year} {2019})},\ \Eprint
  {http://arxiv.org/abs/1811.12928} {arXiv:1811.12928 [astro-ph.CO]}
  \BibitemShut {NoStop}%
\bibitem [{\citenamefont {Spurio~Mancini}\ \emph {et~al.}(2019)\citenamefont
  {Spurio~Mancini}, \citenamefont {K\"ohlinger}, \citenamefont {Joachimi},
  \citenamefont {Pettorino}, \citenamefont {Sch\"afer}, \citenamefont
  {Reischke}, \citenamefont {van Uitert}, \citenamefont {Brieden},
  \citenamefont {Archidiacono},\ and\ \citenamefont
  {Lesgourgues}}]{SpurioMancini:2019rxy}%
  \BibitemOpen
  \bibfield  {author} {\bibinfo {author} {\bibfnamefont {A.}~\bibnamefont
  {Spurio~Mancini}}, \bibinfo {author} {\bibfnamefont {F.}~\bibnamefont
  {K\"ohlinger}}, \bibinfo {author} {\bibfnamefont {B.}~\bibnamefont
  {Joachimi}}, \bibinfo {author} {\bibfnamefont {V.}~\bibnamefont {Pettorino}},
  \bibinfo {author} {\bibfnamefont {B.~M.}\ \bibnamefont {Sch\"afer}}, \bibinfo
  {author} {\bibfnamefont {R.}~\bibnamefont {Reischke}}, \bibinfo {author}
  {\bibfnamefont {E.}~\bibnamefont {van Uitert}}, \bibinfo {author}
  {\bibfnamefont {S.}~\bibnamefont {Brieden}}, \bibinfo {author} {\bibfnamefont
  {M.}~\bibnamefont {Archidiacono}}, \ and\ \bibinfo {author} {\bibfnamefont
  {J.}~\bibnamefont {Lesgourgues}},\ }\href {\doibase 10.1093/mnras/stz2581}
  {\bibfield  {journal} {\bibinfo  {journal} {Mon. Not. Roy. Astron. Soc.}\
  }\textbf {\bibinfo {volume} {490}},\ \bibinfo {pages} {2155} (\bibinfo {year}
  {2019})},\ \Eprint {http://arxiv.org/abs/1901.03686} {arXiv:1901.03686
  [astro-ph.CO]} \BibitemShut {NoStop}%
\bibitem [{\citenamefont {Alam}\ \emph {et~al.}(2021)\citenamefont {Alam} \emph
  {et~al.}}]{eBOSS:2020yzd}%
  \BibitemOpen
  \bibfield  {author} {\bibinfo {author} {\bibfnamefont {S.}~\bibnamefont
  {Alam}} \emph {et~al.} (\bibinfo {collaboration} {eBOSS}),\ }\href {\doibase
  10.1103/PhysRevD.103.083533} {\bibfield  {journal} {\bibinfo  {journal}
  {Phys. Rev. D}\ }\textbf {\bibinfo {volume} {103}},\ \bibinfo {pages}
  {083533} (\bibinfo {year} {2021})},\ \Eprint
  {http://arxiv.org/abs/2007.08991} {arXiv:2007.08991 [astro-ph.CO]}
  \BibitemShut {NoStop}%
\bibitem [{\citenamefont {Tr\"oster}\ \emph {et~al.}(2021)\citenamefont
  {Tr\"oster} \emph {et~al.}}]{KiDS:2020ghu}%
  \BibitemOpen
  \bibfield  {author} {\bibinfo {author} {\bibfnamefont {T.}~\bibnamefont
  {Tr\"oster}} \emph {et~al.} (\bibinfo {collaboration} {KiDS}),\ }\href
  {\doibase 10.1051/0004-6361/202039805} {\bibfield  {journal} {\bibinfo
  {journal} {Astron. Astrophys.}\ }\textbf {\bibinfo {volume} {649}},\ \bibinfo
  {pages} {A88} (\bibinfo {year} {2021})},\ \Eprint
  {http://arxiv.org/abs/2010.16416} {arXiv:2010.16416 [astro-ph.CO]}
  \BibitemShut {NoStop}%
\bibitem [{\citenamefont {Lee}\ \emph {et~al.}(2021)\citenamefont {Lee} \emph
  {et~al.}}]{DES:2021zdr}%
  \BibitemOpen
  \bibfield  {author} {\bibinfo {author} {\bibfnamefont {S.}~\bibnamefont
  {Lee}} \emph {et~al.} (\bibinfo {collaboration} {DES}),\ }\href@noop {} {\
  (\bibinfo {year} {2021})},\ \Eprint {http://arxiv.org/abs/2104.14515}
  {arXiv:2104.14515 [astro-ph.CO]} \BibitemShut {NoStop}%
\bibitem [{\citenamefont {Raveri}(2020)}]{Raveri:2019mxg}%
  \BibitemOpen
  \bibfield  {author} {\bibinfo {author} {\bibfnamefont {M.}~\bibnamefont
  {Raveri}},\ }\href {\doibase 10.1103/PhysRevD.101.083524} {\bibfield
  {journal} {\bibinfo  {journal} {Phys. Rev. D}\ }\textbf {\bibinfo {volume}
  {101}},\ \bibinfo {pages} {083524} (\bibinfo {year} {2020})},\ \Eprint
  {http://arxiv.org/abs/1902.01366} {arXiv:1902.01366 [astro-ph.CO]}
  \BibitemShut {NoStop}%
\bibitem [{\citenamefont {Pogosian}\ \emph {et~al.}(2021)\citenamefont
  {Pogosian}, \citenamefont {Raveri}, \citenamefont {Koyama}, \citenamefont
  {Martinelli}, \citenamefont {Silvestri},\ and\ \citenamefont
  {Zhao}}]{Pogosian:2021mcs}%
  \BibitemOpen
  \bibfield  {author} {\bibinfo {author} {\bibfnamefont {L.}~\bibnamefont
  {Pogosian}}, \bibinfo {author} {\bibfnamefont {M.}~\bibnamefont {Raveri}},
  \bibinfo {author} {\bibfnamefont {K.}~\bibnamefont {Koyama}}, \bibinfo
  {author} {\bibfnamefont {M.}~\bibnamefont {Martinelli}}, \bibinfo {author}
  {\bibfnamefont {A.}~\bibnamefont {Silvestri}}, \ and\ \bibinfo {author}
  {\bibfnamefont {G.-B.}\ \bibnamefont {Zhao}},\ }\href@noop {} {\  (\bibinfo
  {year} {2021})},\ \Eprint {http://arxiv.org/abs/2107.12992} {arXiv:2107.12992
  [astro-ph.CO]} \BibitemShut {NoStop}%
\bibitem [{\citenamefont {Huterer}\ and\ \citenamefont
  {Linder}(2007)}]{Huterer:2006mva}%
  \BibitemOpen
  \bibfield  {author} {\bibinfo {author} {\bibfnamefont {D.}~\bibnamefont
  {Huterer}}\ and\ \bibinfo {author} {\bibfnamefont {E.~V.}\ \bibnamefont
  {Linder}},\ }\href {\doibase 10.1103/PhysRevD.75.023519} {\bibfield
  {journal} {\bibinfo  {journal} {Phys. Rev. D}\ }\textbf {\bibinfo {volume}
  {75}},\ \bibinfo {pages} {023519} (\bibinfo {year} {2007})},\ \Eprint
  {http://arxiv.org/abs/astro-ph/0608681} {arXiv:astro-ph/0608681} \BibitemShut
  {NoStop}%
\bibitem [{\citenamefont {Lin}\ \emph {et~al.}(2019)\citenamefont {Lin},
  \citenamefont {Raveri},\ and\ \citenamefont {Hu}}]{Lin:2018nxe}%
  \BibitemOpen
  \bibfield  {author} {\bibinfo {author} {\bibfnamefont {M.-X.}\ \bibnamefont
  {Lin}}, \bibinfo {author} {\bibfnamefont {M.}~\bibnamefont {Raveri}}, \ and\
  \bibinfo {author} {\bibfnamefont {W.}~\bibnamefont {Hu}},\ }\href {\doibase
  10.1103/PhysRevD.99.043514} {\bibfield  {journal} {\bibinfo  {journal} {Phys.
  Rev. D}\ }\textbf {\bibinfo {volume} {99}},\ \bibinfo {pages} {043514}
  (\bibinfo {year} {2019})},\ \Eprint {http://arxiv.org/abs/1810.02333}
  {arXiv:1810.02333 [astro-ph.CO]} \BibitemShut {NoStop}%
\bibitem [{\citenamefont {Braglia}\ \emph {et~al.}(2021)\citenamefont
  {Braglia}, \citenamefont {Ballardini}, \citenamefont {Finelli},\ and\
  \citenamefont {Koyama}}]{Braglia:2020auw}%
  \BibitemOpen
  \bibfield  {author} {\bibinfo {author} {\bibfnamefont {M.}~\bibnamefont
  {Braglia}}, \bibinfo {author} {\bibfnamefont {M.}~\bibnamefont {Ballardini}},
  \bibinfo {author} {\bibfnamefont {F.}~\bibnamefont {Finelli}}, \ and\
  \bibinfo {author} {\bibfnamefont {K.}~\bibnamefont {Koyama}},\ }\href
  {\doibase 10.1103/PhysRevD.103.043528} {\bibfield  {journal} {\bibinfo
  {journal} {Phys. Rev. D}\ }\textbf {\bibinfo {volume} {103}},\ \bibinfo
  {pages} {043528} (\bibinfo {year} {2021})},\ \Eprint
  {http://arxiv.org/abs/2011.12934} {arXiv:2011.12934 [astro-ph.CO]}
  \BibitemShut {NoStop}%
\bibitem [{\citenamefont {Park}\ \emph {et~al.}(2010)\citenamefont {Park},
  \citenamefont {Zurek},\ and\ \citenamefont {Watson}}]{Park:2010cw}%
  \BibitemOpen
  \bibfield  {author} {\bibinfo {author} {\bibfnamefont {M.}~\bibnamefont
  {Park}}, \bibinfo {author} {\bibfnamefont {K.~M.}\ \bibnamefont {Zurek}}, \
  and\ \bibinfo {author} {\bibfnamefont {S.}~\bibnamefont {Watson}},\ }\href
  {\doibase 10.1103/PhysRevD.81.124008} {\bibfield  {journal} {\bibinfo
  {journal} {Phys. Rev. D}\ }\textbf {\bibinfo {volume} {81}},\ \bibinfo
  {pages} {124008} (\bibinfo {year} {2010})},\ \Eprint
  {http://arxiv.org/abs/1003.1722} {arXiv:1003.1722 [hep-th]} \BibitemShut
  {NoStop}%
\bibitem [{\citenamefont {Gubitosi}\ \emph {et~al.}(2013)\citenamefont
  {Gubitosi}, \citenamefont {Piazza},\ and\ \citenamefont
  {Vernizzi}}]{Gubitosi:2012hu}%
  \BibitemOpen
  \bibfield  {author} {\bibinfo {author} {\bibfnamefont {G.}~\bibnamefont
  {Gubitosi}}, \bibinfo {author} {\bibfnamefont {F.}~\bibnamefont {Piazza}}, \
  and\ \bibinfo {author} {\bibfnamefont {F.}~\bibnamefont {Vernizzi}},\ }\href
  {\doibase 10.1088/1475-7516/2013/02/032} {\bibfield  {journal} {\bibinfo
  {journal} {JCAP}\ }\textbf {\bibinfo {volume} {02}},\ \bibinfo {pages} {032}
  (\bibinfo {year} {2013})},\ \Eprint {http://arxiv.org/abs/1210.0201}
  {arXiv:1210.0201 [hep-th]} \BibitemShut {NoStop}%
\bibitem [{\citenamefont {Bloomfield}\ \emph {et~al.}(2013)\citenamefont
  {Bloomfield}, \citenamefont {Flanagan}, \citenamefont {Park},\ and\
  \citenamefont {Watson}}]{Bloomfield:2012ff}%
  \BibitemOpen
  \bibfield  {author} {\bibinfo {author} {\bibfnamefont {J.~K.}\ \bibnamefont
  {Bloomfield}}, \bibinfo {author} {\bibfnamefont {E.~E.}\ \bibnamefont
  {Flanagan}}, \bibinfo {author} {\bibfnamefont {M.}~\bibnamefont {Park}}, \
  and\ \bibinfo {author} {\bibfnamefont {S.}~\bibnamefont {Watson}},\ }\href
  {\doibase 10.1088/1475-7516/2013/08/010} {\bibfield  {journal} {\bibinfo
  {journal} {JCAP}\ }\textbf {\bibinfo {volume} {08}},\ \bibinfo {pages} {010}
  (\bibinfo {year} {2013})},\ \Eprint {http://arxiv.org/abs/1211.7054}
  {arXiv:1211.7054 [astro-ph.CO]} \BibitemShut {NoStop}%
\bibitem [{\citenamefont {Schneider}\ \emph {et~al.}(2011)\citenamefont
  {Schneider}, \citenamefont {Holm},\ and\ \citenamefont
  {Knox}}]{Schneider2011}%
  \BibitemOpen
  \bibfield  {author} {\bibinfo {author} {\bibfnamefont {M.~D.}\ \bibnamefont
  {Schneider}}, \bibinfo {author} {\bibfnamefont {{\'{O}}.}~\bibnamefont
  {Holm}}, \ and\ \bibinfo {author} {\bibfnamefont {L.}~\bibnamefont {Knox}},\
  }\href {\doibase 10.1088/0004-637X/728/2/137} {\bibfield  {journal} {\bibinfo
   {journal} {Astrophysical Journal}\ }\textbf {\bibinfo {volume} {728}}
  (\bibinfo {year} {2011}),\ 10.1088/0004-637X/728/2/137},\ \Eprint
  {http://arxiv.org/abs/1002.1752} {arXiv:1002.1752} \BibitemShut {NoStop}%
\bibitem [{\citenamefont {Spurio~Mancini}\ \emph {et~al.}(2021)\citenamefont
  {Spurio~Mancini}, \citenamefont {Piras}, \citenamefont {Alsing},
  \citenamefont {Joachimi},\ and\ \citenamefont
  {Hobson}}]{SpurioMancini:2021ppk}%
  \BibitemOpen
  \bibfield  {author} {\bibinfo {author} {\bibfnamefont {A.}~\bibnamefont
  {Spurio~Mancini}}, \bibinfo {author} {\bibfnamefont {D.}~\bibnamefont
  {Piras}}, \bibinfo {author} {\bibfnamefont {J.}~\bibnamefont {Alsing}},
  \bibinfo {author} {\bibfnamefont {B.}~\bibnamefont {Joachimi}}, \ and\
  \bibinfo {author} {\bibfnamefont {M.~P.}\ \bibnamefont {Hobson}},\
  }\href@noop {} {\  (\bibinfo {year} {2021})},\ \Eprint
  {http://arxiv.org/abs/2106.03846} {arXiv:2106.03846 [astro-ph.CO]}
  \BibitemShut {NoStop}%
\bibitem [{\citenamefont {Cheung}\ \emph {et~al.}(2008)\citenamefont {Cheung},
  \citenamefont {Creminelli}, \citenamefont {Fitzpatrick}, \citenamefont
  {Kaplan},\ and\ \citenamefont {Senatore}}]{Cheung:2007st}%
  \BibitemOpen
  \bibfield  {author} {\bibinfo {author} {\bibfnamefont {C.}~\bibnamefont
  {Cheung}}, \bibinfo {author} {\bibfnamefont {P.}~\bibnamefont {Creminelli}},
  \bibinfo {author} {\bibfnamefont {A.~L.}\ \bibnamefont {Fitzpatrick}},
  \bibinfo {author} {\bibfnamefont {J.}~\bibnamefont {Kaplan}}, \ and\ \bibinfo
  {author} {\bibfnamefont {L.}~\bibnamefont {Senatore}},\ }\href {\doibase
  10.1088/1126-6708/2008/03/014} {\bibfield  {journal} {\bibinfo  {journal}
  {JHEP}\ }\textbf {\bibinfo {volume} {03}},\ \bibinfo {pages} {014} (\bibinfo
  {year} {2008})},\ \Eprint {http://arxiv.org/abs/0709.0293} {arXiv:0709.0293
  [hep-th]} \BibitemShut {NoStop}%
\bibitem [{\citenamefont {Linder}\ \emph {et~al.}(2016)\citenamefont {Linder},
  \citenamefont {Seng\"or},\ and\ \citenamefont {Watson}}]{Linder:2015rcz}%
  \BibitemOpen
  \bibfield  {author} {\bibinfo {author} {\bibfnamefont {E.~V.}\ \bibnamefont
  {Linder}}, \bibinfo {author} {\bibfnamefont {G.}~\bibnamefont {Seng\"or}}, \
  and\ \bibinfo {author} {\bibfnamefont {S.}~\bibnamefont {Watson}},\ }\href
  {\doibase 10.1088/1475-7516/2016/05/053} {\bibfield  {journal} {\bibinfo
  {journal} {JCAP}\ }\textbf {\bibinfo {volume} {05}},\ \bibinfo {pages} {053}
  (\bibinfo {year} {2016})},\ \Eprint {http://arxiv.org/abs/1512.06180}
  {arXiv:1512.06180 [astro-ph.CO]} \BibitemShut {NoStop}%
\bibitem [{\citenamefont {Kobayashi}(2019)}]{Kobayashi:2019hrl}%
  \BibitemOpen
  \bibfield  {author} {\bibinfo {author} {\bibfnamefont {T.}~\bibnamefont
  {Kobayashi}},\ }\href {\doibase 10.1088/1361-6633/ab2429} {\bibfield
  {journal} {\bibinfo  {journal} {Rept. Prog. Phys.}\ }\textbf {\bibinfo
  {volume} {82}},\ \bibinfo {pages} {086901} (\bibinfo {year} {2019})},\
  \Eprint {http://arxiv.org/abs/1901.07183} {arXiv:1901.07183 [gr-qc]}
  \BibitemShut {NoStop}%
\bibitem [{\citenamefont {Hu}\ \emph {et~al.}(2014)\citenamefont {Hu},
  \citenamefont {Raveri}, \citenamefont {Frusciante},\ and\ \citenamefont
  {Silvestri}}]{Hu:2014oga}%
  \BibitemOpen
  \bibfield  {author} {\bibinfo {author} {\bibfnamefont {B.}~\bibnamefont
  {Hu}}, \bibinfo {author} {\bibfnamefont {M.}~\bibnamefont {Raveri}}, \bibinfo
  {author} {\bibfnamefont {N.}~\bibnamefont {Frusciante}}, \ and\ \bibinfo
  {author} {\bibfnamefont {A.}~\bibnamefont {Silvestri}},\ }\href@noop {} {\
  (\bibinfo {year} {2014})},\ \Eprint {http://arxiv.org/abs/1405.3590}
  {arXiv:1405.3590 [astro-ph.IM]} \BibitemShut {NoStop}%
\bibitem [{\citenamefont {Kreisch}\ and\ \citenamefont
  {Komatsu}(2018)}]{Kreisch:2017uet}%
  \BibitemOpen
  \bibfield  {author} {\bibinfo {author} {\bibfnamefont {C.~D.}\ \bibnamefont
  {Kreisch}}\ and\ \bibinfo {author} {\bibfnamefont {E.}~\bibnamefont
  {Komatsu}},\ }\href {\doibase 10.1088/1475-7516/2018/12/030} {\bibfield
  {journal} {\bibinfo  {journal} {JCAP}\ }\textbf {\bibinfo {volume} {12}},\
  \bibinfo {pages} {030} (\bibinfo {year} {2018})},\ \Eprint
  {http://arxiv.org/abs/1712.02710} {arXiv:1712.02710 [astro-ph.CO]}
  \BibitemShut {NoStop}%
\bibitem [{\citenamefont {de~Rham}\ and\ \citenamefont
  {Melville}(2018)}]{deRham:2018red}%
  \BibitemOpen
  \bibfield  {author} {\bibinfo {author} {\bibfnamefont {C.}~\bibnamefont
  {de~Rham}}\ and\ \bibinfo {author} {\bibfnamefont {S.}~\bibnamefont
  {Melville}},\ }\href {\doibase 10.1103/PhysRevLett.121.221101} {\bibfield
  {journal} {\bibinfo  {journal} {Phys. Rev. Lett.}\ }\textbf {\bibinfo
  {volume} {121}},\ \bibinfo {pages} {221101} (\bibinfo {year} {2018})},\
  \Eprint {http://arxiv.org/abs/1806.09417} {arXiv:1806.09417 [hep-th]}
  \BibitemShut {NoStop}%
\bibitem [{\citenamefont {Amendola}\ \emph {et~al.}(2018)\citenamefont
  {Amendola}, \citenamefont {Bettoni}, \citenamefont {Dom\`enech},\ and\
  \citenamefont {Gomes}}]{Amendola:2018ltt}%
  \BibitemOpen
  \bibfield  {author} {\bibinfo {author} {\bibfnamefont {L.}~\bibnamefont
  {Amendola}}, \bibinfo {author} {\bibfnamefont {D.}~\bibnamefont {Bettoni}},
  \bibinfo {author} {\bibfnamefont {G.}~\bibnamefont {Dom\`enech}}, \ and\
  \bibinfo {author} {\bibfnamefont {A.~R.}\ \bibnamefont {Gomes}},\ }\href
  {\doibase 10.1088/1475-7516/2018/06/029} {\bibfield  {journal} {\bibinfo
  {journal} {JCAP}\ }\textbf {\bibinfo {volume} {06}},\ \bibinfo {pages} {029}
  (\bibinfo {year} {2018})},\ \Eprint {http://arxiv.org/abs/1803.06368}
  {arXiv:1803.06368 [gr-qc]} \BibitemShut {NoStop}%
\bibitem [{\citenamefont {Battye}\ \emph {et~al.}(2018)\citenamefont {Battye},
  \citenamefont {Pace},\ and\ \citenamefont {Trinh}}]{Battye:2018ssx}%
  \BibitemOpen
  \bibfield  {author} {\bibinfo {author} {\bibfnamefont {R.~A.}\ \bibnamefont
  {Battye}}, \bibinfo {author} {\bibfnamefont {F.}~\bibnamefont {Pace}}, \ and\
  \bibinfo {author} {\bibfnamefont {D.}~\bibnamefont {Trinh}},\ }\href
  {\doibase 10.1103/PhysRevD.98.023504} {\bibfield  {journal} {\bibinfo
  {journal} {Phys. Rev. D}\ }\textbf {\bibinfo {volume} {98}},\ \bibinfo
  {pages} {023504} (\bibinfo {year} {2018})},\ \Eprint
  {http://arxiv.org/abs/1802.09447} {arXiv:1802.09447 [astro-ph.CO]}
  \BibitemShut {NoStop}%
\bibitem [{\citenamefont {Aghanim}\ \emph
  {et~al.}(2020{\natexlab{b}})\citenamefont {Aghanim} \emph
  {et~al.}}]{Planck:2018vyg}%
  \BibitemOpen
  \bibfield  {author} {\bibinfo {author} {\bibfnamefont {N.}~\bibnamefont
  {Aghanim}} \emph {et~al.} (\bibinfo {collaboration} {Planck}),\ }\href
  {\doibase 10.1051/0004-6361/201833910} {\bibfield  {journal} {\bibinfo
  {journal} {Astron. Astrophys.}\ }\textbf {\bibinfo {volume} {641}},\ \bibinfo
  {pages} {A6} (\bibinfo {year} {2020}{\natexlab{b}})},\ \bibinfo {note}
  {[Erratum: Astron.Astrophys. 652, C4 (2021)]},\ \Eprint
  {http://arxiv.org/abs/1807.06209} {arXiv:1807.06209 [astro-ph.CO]}
  \BibitemShut {NoStop}%
\bibitem [{\citenamefont {Heitmann}\ \emph {et~al.}(2009)\citenamefont
  {Heitmann}, \citenamefont {Higdon}, \citenamefont {White}, \citenamefont
  {Habib}, \citenamefont {Williams}, \citenamefont {Lawrence},\ and\
  \citenamefont {Wagner}}]{Heitmann2009}%
  \BibitemOpen
  \bibfield  {author} {\bibinfo {author} {\bibfnamefont {K.}~\bibnamefont
  {Heitmann}}, \bibinfo {author} {\bibfnamefont {D.}~\bibnamefont {Higdon}},
  \bibinfo {author} {\bibfnamefont {M.}~\bibnamefont {White}}, \bibinfo
  {author} {\bibfnamefont {S.}~\bibnamefont {Habib}}, \bibinfo {author}
  {\bibfnamefont {B.~J.}\ \bibnamefont {Williams}}, \bibinfo {author}
  {\bibfnamefont {E.}~\bibnamefont {Lawrence}}, \ and\ \bibinfo {author}
  {\bibfnamefont {C.}~\bibnamefont {Wagner}},\ }\href {\doibase
  10.1088/0004-637X/705/1/156} {\bibfield  {journal} {\bibinfo  {journal}
  {Astrophysical Journal}\ }\textbf {\bibinfo {volume} {705}},\ \bibinfo
  {pages} {156} (\bibinfo {year} {2009})},\ \Eprint
  {http://arxiv.org/abs/arXiv:0902.0429v2} {arXiv:arXiv:0902.0429v2}
  \BibitemShut {NoStop}%
\bibitem [{\citenamefont {Scolnic}\ \emph {et~al.}(2018)\citenamefont {Scolnic}
  \emph {et~al.}}]{Scolnic:2017caz}%
  \BibitemOpen
  \bibfield  {author} {\bibinfo {author} {\bibfnamefont {D.~M.}\ \bibnamefont
  {Scolnic}} \emph {et~al.},\ }\href {\doibase 10.3847/1538-4357/aab9bb}
  {\bibfield  {journal} {\bibinfo  {journal} {Astrophys. J.}\ }\textbf
  {\bibinfo {volume} {859}},\ \bibinfo {pages} {101} (\bibinfo {year}
  {2018})},\ \Eprint {http://arxiv.org/abs/1710.00845} {arXiv:1710.00845
  [astro-ph.CO]} \BibitemShut {NoStop}%
\bibitem [{\citenamefont {Aghamousa}\ \emph {et~al.}(2016)\citenamefont
  {Aghamousa} \emph {et~al.}}]{DESI:2016fyo}%
  \BibitemOpen
  \bibfield  {author} {\bibinfo {author} {\bibfnamefont {A.}~\bibnamefont
  {Aghamousa}} \emph {et~al.} (\bibinfo {collaboration} {DESI}),\ }\href@noop
  {} {\  (\bibinfo {year} {2016})},\ \Eprint {http://arxiv.org/abs/1611.00036}
  {arXiv:1611.00036 [astro-ph.IM]} \BibitemShut {NoStop}%
\bibitem [{\citenamefont {Li}\ \emph {et~al.}(2018)\citenamefont {Li},
  \citenamefont {Weaverdyck}, \citenamefont {Adhikari}, \citenamefont
  {Huterer}, \citenamefont {Muir},\ and\ \citenamefont {Wu}}]{Li:2018epc}%
  \BibitemOpen
  \bibfield  {author} {\bibinfo {author} {\bibfnamefont {X.}~\bibnamefont
  {Li}}, \bibinfo {author} {\bibfnamefont {N.}~\bibnamefont {Weaverdyck}},
  \bibinfo {author} {\bibfnamefont {S.}~\bibnamefont {Adhikari}}, \bibinfo
  {author} {\bibfnamefont {D.}~\bibnamefont {Huterer}}, \bibinfo {author}
  {\bibfnamefont {J.}~\bibnamefont {Muir}}, \ and\ \bibinfo {author}
  {\bibfnamefont {H.-Y.}\ \bibnamefont {Wu}},\ }\href {\doibase
  10.3847/1538-4357/aacaf7} {\bibfield  {journal} {\bibinfo  {journal}
  {Astrophys. J.}\ }\textbf {\bibinfo {volume} {862}},\ \bibinfo {pages} {137}
  (\bibinfo {year} {2018})},\ \Eprint {http://arxiv.org/abs/1806.02515}
  {arXiv:1806.02515 [astro-ph.CO]} \BibitemShut {NoStop}%
\bibitem [{\citenamefont {Hounsell}\ \emph {et~al.}(2018)\citenamefont
  {Hounsell} \emph {et~al.}}]{Hounsell:2017ejq}%
  \BibitemOpen
  \bibfield  {author} {\bibinfo {author} {\bibfnamefont {R.}~\bibnamefont
  {Hounsell}} \emph {et~al.},\ }\href {\doibase 10.3847/1538-4357/aac08b}
  {\bibfield  {journal} {\bibinfo  {journal} {Astrophys. J.}\ }\textbf
  {\bibinfo {volume} {867}},\ \bibinfo {pages} {23} (\bibinfo {year} {2018})},\
  \Eprint {http://arxiv.org/abs/1702.01747} {arXiv:1702.01747 [astro-ph.IM]}
  \BibitemShut {NoStop}%
\bibitem [{\citenamefont {Huterer}\ and\ \citenamefont
  {Starkman}(2003)}]{Huterer:2002hy}%
  \BibitemOpen
  \bibfield  {author} {\bibinfo {author} {\bibfnamefont {D.}~\bibnamefont
  {Huterer}}\ and\ \bibinfo {author} {\bibfnamefont {G.}~\bibnamefont
  {Starkman}},\ }\href {\doibase 10.1103/PhysRevLett.90.031301} {\bibfield
  {journal} {\bibinfo  {journal} {Phys. Rev. Lett.}\ }\textbf {\bibinfo
  {volume} {90}},\ \bibinfo {pages} {031301} (\bibinfo {year} {2003})},\
  \Eprint {http://arxiv.org/abs/astro-ph/0207517} {arXiv:astro-ph/0207517}
  \BibitemShut {NoStop}%
\bibitem [{\citenamefont {Linder}(2006)}]{Linder:2006xb}%
  \BibitemOpen
  \bibfield  {author} {\bibinfo {author} {\bibfnamefont {E.~V.}\ \bibnamefont
  {Linder}},\ }\href {\doibase 10.1016/j.astropartphys.2006.05.004} {\bibfield
  {journal} {\bibinfo  {journal} {Astropart. Phys.}\ }\textbf {\bibinfo
  {volume} {26}},\ \bibinfo {pages} {102} (\bibinfo {year} {2006})},\ \Eprint
  {http://arxiv.org/abs/astro-ph/0604280} {arXiv:astro-ph/0604280} \BibitemShut
  {NoStop}%
\bibitem [{\citenamefont {Linder}(2007)}]{Linder:2007ka}%
  \BibitemOpen
  \bibfield  {author} {\bibinfo {author} {\bibfnamefont {E.~V.}\ \bibnamefont
  {Linder}},\ }\href@noop {} {\  (\bibinfo {year} {2007})},\ \Eprint
  {http://arxiv.org/abs/0708.0024} {arXiv:0708.0024 [astro-ph]} \BibitemShut
  {NoStop}%
\end{thebibliography}%
\end{document}